\documentclass[prb,aps,amsmath,amssymb,amsfonts,floatfix,groupedaddress,showpacs,twocolumn]{revtex4-1}
\usepackage{longtable}
\usepackage{graphicx}
\usepackage{dcolumn}
\usepackage{epsfig}
\usepackage[dvips]{color}
\usepackage{dcolumn}
\usepackage{nicefrac}
\usepackage{float}
\usepackage{tabularx}
\newcolumntype{L}[1]{>{\raggedright\arraybackslash}p{#1}} 
\newcolumntype{C}[1]{>{\centering\arraybackslash}p{#1}} 
\newcolumntype{R}[1]{>{\raggedleft\arraybackslash}p{#1}} 

\newcommand{\mub}{\mu_{\mbox{\tiny{B}}}}

\newcommand{\bk}{\mathbf{k}}

\newcommand{\bH}{\mathbf{H}}
\newcommand{\ham}{{\cal H}}

\def\beq{\begin{equation}}
\def\eeq{\end{equation}}
\def\be{\begin{equation}}
\def\ee{\end{equation}}

\def\t{\mbox{tr}\,}
\def\cG0{{\cal G}_0}
\def\cG{{\cal G}}

\def\efat{\mbox{\boldmath$\varepsilon$}}
\def\a{\alpha}

\def\uc2{$U_{c2}$}
\def\uc1{$U_{c1}$}
\def\bavs3{BaVS$_3$}

\def\t2g{$t_{2g}$}

\def\a1g{$a_{1g}$}

\usepackage[utf8]{inputenc}
\usepackage[T1]{fontenc}
\newcommand{\ii}{\mathrm{i}}
\newcommand{\bL}{\mathbf{L}}
\newcommand{\bS}{\mathbf{S}}
\newcommand{\bJ}{\mathbf{J}}

\begin{document}

\title{Multi-orbital physics in Fermi liquids prone to magnetic order}
\author{Malte Behrmann}
\affiliation{I. Institut f{\"u}r Theoretische Physik,
Universit{\"a}t Hamburg, D-20355 Hamburg, Germany}
\author{Christoph Piefke}
\affiliation{I. Institut f{\"u}r Theoretische Physik,
Universit{\"a}t Hamburg, D-20355 Hamburg, Germany}
\author{Frank Lechermann}
\affiliation{I. Institut f{\"u}r Theoretische Physik, 
Universit{\"a}t Hamburg, D-20355 Hamburg, Germany}

\begin{abstract}
The interplay of spin-orbit-coupling and strong electronic correlations is
studied for the single-layer and the bilayer compound of the strontium ruthenate
Ruddlesden-Popper series by a combination of first-principles band-structure theory
with mean-field rotationally invariant slave bosons. At equilibrium strongly
renormalized (spin-orbit-split) quasiparticle bands are traced and a thorough
description of the low-energy regime for the nearly ferromagnetic bilayer system in 
accordance with experimental data is presented. The metamagnetic response of 
Sr$_3$Ru$_2$O$_7$ in finite magnetic field $H$ is verified and a detailed analysis of 
the underlying correlated electronic structure provided. Intriguing multi-orbital 
physics on both local and itinerant level, such as e.g. competing paramagnetic and
diamagnetic contributions, is observed with important differences depending on the 
magnetic-field angle $\theta$ with the crystallographic $c$ axis.
\end{abstract}

\pacs{71.27.+a, 71.18.+y, 75.10.Lp, 75.70.Tj, 75.47.Lx}
\maketitle

\section{Introduction}
Ternary ruthenium-based oxide compounds condense in a rich variety of different
crystal-structure types with associated rather delicate electronic and magnetic 
properties.~\cite{cav04} Within the perovskite-like Ruddlesden-Popper series of the 
ruthenates A$_{n+1}$Ru$_n$O$_{3n+1}$ (A=Sr,Ca), where $n$ labels the layers of 
corner-sharing RuO$_6$ octahedra (separated by SrO or CaO rocksalt layers), the
interacting electrons pose a specifically challenging problem. In that family of 
seemingly rather similar compounds, intriguing competitions between Fermi-liquid, 
Mott-insulating and superconducting behavior occur in conjunction with particular 
complex magnetic response. For instance, albeit Ca and Sr are isovalent, the 
respective single-layer compounds A$_2$RuO$_4$ exhibit drastic different 
phenomenology, since Sr$_2$RuO$_4$ displays unconventional superconductivity 
below $T_{\rm c}$$\sim$1.5 K~\cite{mae94,ric95,mac03}, while Ca$_2$RuO$_4$ becomes an 
antiferromagnetic (AFM) Mott insulator below 
$T_{\rm MI}$$\sim$77 K.~\cite{ale99,gor10} As an important generic distinction, 
compared to the Sr compounds the Ca subclass in the series exhibits stronger 
distortions and deviations from an ideal high crystal symmetry. 

\begin{figure}[t]
\centering
\includegraphics*[width=8cm]{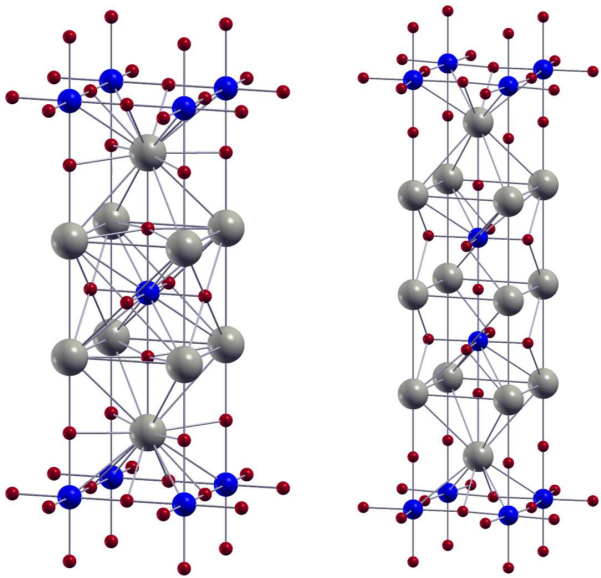}%
\caption{(Color online) Crystal structures of Sr$_2$RuO$_4$ (left) and
Sr$_3$Ru$_2$O$_7$ (right). Large grey: Sr, blue (dark): Ru and small red (dark): O. 
\label{fig:214-327}}
\end{figure}

A prominent aspect of the intricate physics in the overall metallic strontium
ruthenates is the onset of ferromagnetism with $n$. Concerning the series end members,
the perovskite SrRuO$_3$ ($n$$\rightarrow$$\infty$) is ferromagnetic (FM) below 
$T_{\rm C}$$\sim$165 K and tetragonal Sr$_2$RuO$_4$ ($n$=1) is paramagnetic (PM) at 
ambient temperature $T$, but shows FM tendencies as well as incommensurate
spin fluctuations at ${\bf q}$=$(\pm 0.6\pi/a,\pm 0.6\pi/a,0)$.~\cite{maz99,sid99} 
The ($n$=3) Sr$_4$Ru$_3$O$_{10}$ and the ($n$=2) Sr$_3$Ru$_2$O$_7$ compounds are
both orthorhombic, but whereas the former is verified FM,~\cite{cra02} the latter
is still PM down to low temperatures. However the bilayer system appears to be located 
rather close to the transition towards FM order,~\cite{ike00} with puzzling 
metamagnetic (MM) behavior in applied field below $T_{\rm MM}$$\sim$ 1 K 
(see Ref.~\onlinecite{mac12} for a recent review). As
revealed from de Haas-van Alphen (dHvA),~\cite{mac96} angle-resolved photoemission 
spectroscopy (ARPES),~\cite{yok96,dam00,she01,tam08} optics~\cite{mir08} and 
resistivity~\cite{ike00} measurements, the $n$=1,2 compounds 
(see Fig.~\ref{fig:214-327}) both belong to cases of 
quasi-twodimensional (2D) electron systems, i.e., show a strong anisotropy between 
transport in the $ab$-plane and along the $c$-axis of the crystal structure. 
Signatures of strong electronic correlations are compelling for these layered
ruthenates, e.g. from large mass renormalizations,~\cite{mae97,ike00,qu08} and 
originate from the less-screened Coulomb interactions within the Ru$(4d)$ shell. 
Nominally four electrons occupy this $l$=2 manifold, i.e. an Ru$^{4+}$ oxidation 
state may be assumed.
The metamagnetism with applied field $H$ in Sr$_3$Ru$_2$O$_7$ is well documented
by a large slope $\frac{\partial M}{\partial H}|_{H_{\rm MM}}$ around 
$H_{\rm MM}$=5.5(7.7)T for $H$$||$$ab$($c$).~\cite{perr01,ros09} Furthermore this 
MM region may be associated with being in the neighborhood of a quantum-critial point 
that can be approached via tuning the polar angle $\theta$ between magnetic field and 
the $c$-axis.~\cite{gri03,geg06} Reachable within fields $H$$<$10 T, the MM phenomena 
are acting on a very low energy scale of the order of at most a few meV. In this 
respect, the Fermi-liquid regime in vanishing field exists below 10-15 K, however can 
be driven to zero temperature with applied field.~\cite{gri01}

In this work we present a theoretical investigation of the $n$=1,2 compounds 
in the normal state of the low-temperature regime with an emphasis on the intriguing
physics of the bilayer system in applied magnetic field. The study is based on the 
combination of a first-principles band-structure approach, spin-orbit-interaction 
treatment in the Russell-Saunders limit and mean-field many-body theory. The peculiar 
low-energy physics of the layered ruthenates ask for a very detailed examination of 
the single- and many-particle terms in the Hamiltonian in order to capture the 
important processes that drive the physics of these systems.~\cite{ric95,cuo98,maz99,lie00,spa01,pav06,pch07,hav08,mra11,roz11,dei11,sin01,bin04,gri04,kee05,pue07,tam08,ber09,rag09,leewu09_2,lee10,lee10_2,fis10,pue10,pie11,mal11,mra11} Besides the thorough description 
of the crystal bonding which leads to a multi-orbital based 
band manifold at the Fermi level, it was shown~\cite{pav06,hav08,roz11} that 
additionally spin-orbit effects play a vital role in the low-energy regime. 
Focussing on the many-body part, e.g. the relevance of the Hund's coupling $J_{\rm H}$ in
addition to the larger Hubbard $U$ was elucidated in several works.~\cite{cuo98,spa01,mra11} 
Much theoretical effort has also been devoted to the 
description of the MM phenomena in Sr$_3$Ru$_2$O$_7$, either based on effective
single-band modelings~\cite{bin04,gri04,kee05,pue07,ber09,fis10} or with including 
multi-band degrees of freedom.~\cite{rag09,leewu09_2,lee10,lee10_2,pue10} 
The existing model studies are able to account for the principle appearance of 
metamagnetism, often accompanied by nematic order,~\cite{kee05} i.e. broken 
rotational symmetry. Important ingredients for the MM behavior are van-Hove 
singularities close to the Fermi surface, already revealed in the single-band 
approaches.~\cite{bin04} More sophisticated multi-orbital investigations have 
been employed to discriminate between the importance of $d_{xz}$, $d_{yz}$ (formally
quasi-1D like dispersions) and $d_{xy}$ (formally quasi-2D like dispersions) 
orbital degrees of freedom together with spin-orbit coupling (SOC). However often 
the broken fourfold symmetry in Sr$_3$Ru$_2$O$_7$ is neglected, when modeling the 
electronic states. In addition, most theoretical works treat the many-body
interactions in the Hartree-Fock approximation, not allowing for explicit 
self-consistent renormalizations and ill-defined for the metallic regime.

A combination of the local density approximation (LDA) to density functional
theory (DFT) with slave-boson~\cite{bar76,rea83,col84,kot86} theory in a
rotationally invariant representation~\cite{li89,lec07} was utilized in 
Ref.~\onlinecite{pie11} in order to account for renormalized quasiparticle (QP) 
behavior. Here an extension of that work is provided by including the effect of
the spin-orbit coupling on the Fermi-liquid regime of the layered ruthenates
including applied magnetic fields. In the case of Sr$_3$Ru$_2$O$_7$ the main focus is 
on a many-body modeling that starts from the realistic low-symmetry band 
structure. Because of the fact that the MM problem involves very low-energy scales 
in an underlying low-symmetry lattice, we believe that the bilayer system serves as a
challenging test case for the reliability and accuracy of current extended-LDA 
approaches.

\section{Theoretical Approach}
The LDA part of this work is performed using an implementation~\cite{mbpp_code} of 
the highly-accurate mixed-basis pseudopotential (MBPP) technique,~\cite{lou79}
employing normconserving pseudopotentials~\cite{van85} and an efficient combined 
basis consisting of plane waves and additional localized orbitals. For the 
inclusion of spin-orbit coupling and to treat local many-body interactions, the 
realistic Hamiltonian 
\begin{equation}                                     
\ham=\sum_{\bk ijmm'\sigma}\varepsilon^{{\rm (ks)}\prime}_{\bk ijmm'}\,       
d^\dagger_{\bk im\sigma}d^{\hfill}_{\bk jm'\sigma}+
\sum_\alpha \ham^{\rm (loc)}_\alpha\;,   
\label{eq:fullham}                          
\end{equation}
is used, where $\bk$ denotes the wave vector, $\alpha$ numbers the unit cells
with $i,j$ marking the Ru ions within, the spin-projection is given by 
$\sigma$=$\uparrow,\downarrow$ and 
$d^{(\dagger)}$ annihilates (creates) electrons in the $t_{2g}$-like Wannier 
orbitals $m$, $m'$. The Kohn-Sham dispersion $\efat_{\bk}$ in the latter basis is 
here obtained from a maximally-localized Wannier-function (MLWF)
construction~\cite{mar97,sou01} based on the band structure revealed from the MBPP
calculation. The prime in (\ref{eq:fullham}) indicates that
strictly local (on-site) contributions are excluded in the $k$-dependent dispersion
(see eq. (\ref{eq:cf})). More details on the determination of the dispersive
part in the case of Sr$_3$Ru$_2$O$_7$, where the band Hamiltonian amounts to an 
12$\times$12 matrix due to the fact that the primitve unit cell encloses four Ru 
ions, is provided in Ref.~\onlinecite{pie11}. For Sr$_2$RuO$_4$ the primitive
unit cell contains only one Ru ion and the Kohn-Sham problem asks for the
diagonalization of a 3$\times$3 matrix. The unit-cell Hamiltonian 
$\ham^{\rm (loc)}_\alpha$ decomposes into the four terms, reading
\begin{equation}
\ham_\alpha^{\rm(loc)}=\ham_\alpha^{\rm(cf)}+\ham_\alpha^{\rm(soc)}+
\ham_\alpha^{\rm(zm)}+\ham_\alpha^{\rm(int)}
\label{eq:hamloc}
\end{equation}
and is evalutated in each particle sector of an effective $t_{2g}$ problem. 
The first contribution includes the on-site crystal-field through
\begin{equation}
\ham_\alpha^{\rm(cf)}=\sum_{imm'\sigma}\varepsilon_{imm'}^{\rm (ks), loc}\,
d^\dagger_{im\sigma}d_{im'\sigma}\;,
\label{eq:cf}
\end{equation}
with $\varepsilon_{imm'}^{\rm (ks), loc}$=$1/N_{\bk}\sum_{\bk}\varepsilon_{\bk iimm'}^{\rm (ks)}$ computed from the complete Kohn-Sham dispersion. We continue with
the spin-orbit interaction $\ham_\alpha^{\rm(soc)}$ of Russell-Saunders type 
(or $LS$ coupling scheme) on each of the rather light Ru ions. Summing over the
individual ion contributions leads to
\begin{equation}
\ham_\alpha^{\rm(soc)} = \lambda \sum_i \mathbf{L}_i \cdot\mathbf{S}_i = 
\frac{\lambda}{2} \sum_i \left( \mathbf{J}_i^2 - \mathbf{L}_i^2 -\mathbf{S}_i^2 
\right)\quad ,
\end{equation}
where $\lambda$ is the spin-orbit coupling parameter, $\mathbf{L}_i$ the total orbital 
momentum operator, $\mathbf{S}_i$ the total spin operator and 
$\mathbf{J}_i$=$\bL_i$$+$$\bS_i$ the total angular momentum operator for each Ru ion,
respectively. Note that in the Russell-Saunders approximation these
three operators are true many-particle operators, given by the sum over the respective
operators for each individual electron $p$ within the $t_{2g}$ shell, e.g. 
$\mathbf{L}_i=\sum_p \mathbf{L}_{ip}$ for the total orbital momentum operator.
For the $t_{2g}$ orbitals a valid choice for the matrix elements 
$\left< m \left | \mathbf{L}_{ip} \right | m' \right>$ reads component-resolved~\cite{lee10}
\begin{equation}
L^x = \left (\begin{matrix}
  0 & 0 & \ii \\
  0 & 0 & 0 \\
  -\ii & 0 & 0
 \end{matrix}
\right ), \: L^y = \left (\begin{matrix} 
  0 & 0 & 0 \\    
  0 & 0 & -\ii \\ 
  0 & \ii & 0   
\end{matrix}        
\right ),
 \: L^z = \left (\begin{matrix}   
  0 & -\ii & 0 \\      
  \ii & 0 & 0 \\    
  0 & 0 & 0      
 \end{matrix}\right )\,.
\end{equation}
Due to the cubic crystal-field terms in the ruthenates, this representation is obtained by 
truncating the appropriate matrix elements of a full $4d$ shell based on cubic harmonics to 
a pure $t_{2g}$ shell including the states $4d_{xz}$, $4d_{yz}$ and $4d_{xy}$. 
Although there is some minor inter-mixing with the $e_g$ states at low-energy for 
Sr$_3$Ru$_2$O$_7$ (see section~\ref{corrstruc}), that approximation proves to
be adequate on the present level of the investigation.
   
The third contribution to eq. (\ref{eq:hamloc}) describes the local interaction of 
the Zeeman type with a magnetic field $\mathbf{H}$ and can be written as
\begin{equation}
\ham_\alpha^{\mathrm(zm)}= \mub \sum_i \left (\bL_i + 2\bS_i \right)\cdot\bH\quad.
\end{equation} 
Notice that in the weak-coupling regime $|\bH|$$\ll$$\lambda$ considered in 
our calculations, the spin-orbit coupling dominates the magnetic-field interaction.
Thus eigenvalues of $L_z$ and $S_z$ are no good quantum numbers of the system, but
$\{J^2,J_z\}$ are now commuting with ${\cal H}^{\rm (loc)}_{\alpha}$. Therefore we have to 
perform the projection of ${\bf L}_i$, ${\bf S}_i$ onto ${\bf J}_i$ according to the 
Wigner-Eckart theorem using the operators' common eigenspace representation.
The term $\ham_\alpha^{\rm(zm)}$ has then the following form:
\begin{eqnarray}
\ham_\alpha^{\mathrm(zm)} &=& 
\mub \sum_i \left (\frac{\left < \bL_i \cdot \bJ_i \right >_{LSJ}}
{\left <J_i^2 \right >_{LSJ}} + 2 \frac{\left < \bS_i \cdot \bJ_i \right >_{LSJ}}
{\left <J_i^2 \right >_{LSJ}}  \right) \bJ_i \cdot \bH \nonumber\\
&=&\mub \sum_i \left (\frac{3}{2} + \frac{\left < S_i^2 \right >_{LSJ} - 
\left < L_i^2 \right >_{LSJ}}{2 \left < J_i^2 \right >_{LSJ}} \right )
\bJ_i \cdot \bH\nonumber\\ 
&=&\mub \sum_i \left (\frac{3}{2} + \frac{S(S+1) - L(L+1)}
{2 J(J+1)} \right )
\bJ_i \cdot \bH\nonumber\\ 
&\equiv& \mub \sum_i g_i(LSJ)\, \bJ_i \cdot \bH\;,
\label{eq:zeeman}
\end{eqnarray}
where the notation $\left < \dots \right >_{LSJ}$ indicates the expectation value in
that eigenspace defined by the eigenvalues of the operators $L^2$, $S^2$ and $J^2$, 
labelled by the quantum numbers $L$, $S$, and $J$, respectively.~\cite{cot08} 
Notably, the object $g_i(LSJ)$ is the generic matrix representation of the Land\'{e} factor 
(or $g$-factor) in that eigenspace. After that eigenspace computation 
one has to transform $\ham_\alpha^{\rm(zm)}$ back into the original Hilbert 
space, where all other parts of the local Hamiltonian were derived. Note that in this
way the $g$-factor is calculated seperately for each considered state. 

Last but not least, eq.~(\ref{eq:hamloc}) includes the electron-electron interaction 
$\ham_\alpha^{\rm(int)}$ provided by a multi-orbital Hubbard model, which reads
\begin{eqnarray}      
\ham^{\rm (int)}_\alpha&=&U\sum_{im} n_{im\uparrow}n_{im\downarrow}    
+\frac 12 \sum \limits _{i,m \ne m',\sigma}              
\Big\{U' \, n_{im \sigma} n_{im' \bar \sigma}\nonumber\\     
&&+ U'' \,n_{im \sigma}n_{im' \sigma}+           
J_{\rm H}\, d^\dagger_{im \sigma} d^\dagger_{im' \bar\sigma} 
d^{\hfill}_{im \bar \sigma} d^{\hfill}_{im' \sigma}\nonumber\\    
&&+\left. J_{\rm H}\,d^\dagger_{im \sigma} d^\dagger_{im \bar \sigma}  
 d^{\hfill}_{im' \bar \sigma} d^{\hfill}_{im' \sigma}\right\}         
\label{eq:hubbardham}          
\end{eqnarray} 
with $n$=$d^\dagger d$. In 
eq. (\ref{eq:hubbardham}) the first term marks the intra-orbital Coulomb interaction 
with Hubbard $U$ and the second term provides inter-orbital Hund's rule corrected 
interaction with $U'$=$U$$-$$J_{\rm H}$ and $U''$=$U$$-$$2J_{\rm H}$ for unequal and equal 
spin projections $\sigma$, respectively. The last two parts account for 
spin-flip and pair-hopping processes, vital to enforce the rotational invariance.
These terms are especially important concerning the magnetic response of an
interacting system.~\cite{li89}

The complete Hamiltonian~(\ref{eq:fullham}) embodies three interaction parameters,
namely the spin-orbit interaction $\lambda$, the Hubbard $U$ and the Hund's exchange
$J_{\rm H}$. For both systems the value $\lambda$=0.09 eV, as obtained from LDA calculations 
for Sr$_2$RuO$_4$ by Haverkort {\sl et al.}~\cite{hav08}, is used for the SOC. 
Concerning the Coulomb interactions, previous works~\cite{cuo98,lie00,pch07,mra11,mal11} 
located the Hubbard $U$ for the layered ruthenates in the region 1.5$-$3.1 eV. Here 
we choose the moderate value $U$=2 eV. With including SOC, that order
of magnitude is sufficient to account for the key renormalization effects at low 
energy. The Hund's exchange is fixed to $J_{\rm H}$=0.35 eV~\cite{cuo98,lie00,oka04,mra11} 
throughout this work.

Our interacting ruthenate problem is solved via the rotationally invariant 
slave-boson (RISB) formalism~\cite{li89,lec07} in the saddle-point approximation. 
It amounts to a decomposition of the electron's QP (fermionic $f_{\nu\sigma}$) and 
high-energy excitations (taken care of by the set of slave bosons $\{\phi\}$) on the 
operator level through $\underline{d}_{\nu\sigma}$=$\hat{R}[\{\phi\}]^{\sigma\sigma'}_{\nu\nu'}f_{\nu'\sigma'}$, where $\nu$ is a generic orbital/site index. Additional 
constraints for the normalization and to match the fermionic and bosonic 
contents are enforced on the mean-field level. The RISB electronic self-energy 
$\Sigma(\omega)$ at saddle-point is local and incorporates terms linear in frequency 
as well as static renormalizations. It is thus given by
\begin{eqnarray}
\mathbf{\Sigma}(\omega)&=&\omega\left(1-\mathbf{Z}^{-1}\right)
+{\bf \Sigma}^{\rm stat}\,\,,\label{eq:Sigma_physical1}\\
\mbox{with}&& {\bf \Sigma}^{\rm stat}=
[\mathbf{R}^\dagger]^{-1}\mathbf{\Lambda}\mathbf{R}^{-1}-\efat^{\rm (ks), loc}\,\,,
\label{eq:Sigma_physical2}
\end{eqnarray}
whereby $\mathbf{Z}$ is the QP-weight matrix and $\mathbf{\Lambda}$ describes
the matrix of Lagrange multipliers for the enforcement of the constraints.
Expectation values of any given local operator ${\cal O}$ may be computed via the
slave bosons according to
\begin{equation}
\langle {\cal O}\rangle =\sum_{AB}\langle A|{\cal O}|B\rangle
\sum_q\phi_{Aq}^*\phi_{Bq}^{\hfill}\,\,,
\label{eq:locex}
\end{equation}
with $A,B$ denoting atomic states and $q$ as the QP index. For more details see 
Ref.~\onlinecite{lec07}. In the present scope the method may also
be interpreted as a simplified approach to solve the dynamical mean-field theory 
(DMFT) equations (see  e.g. Ref.~\onlinecite{geo96} for a review), compared to e.g. 
more elaborate quantum Monte-Carlo (QMC) techniques. Since we are interested in the 
low-energy physics of the layered ruthenates at rather small temperatures (where
QMC usually becomes very challenging), this approach is thus well suited to access 
the Fermi-liquid regime including its extension to magnetically ordered phases.

\section{ Correlated electronic structure of the $n$=1,2 compounds\label{corrstruc}}
\begin{figure}[b]
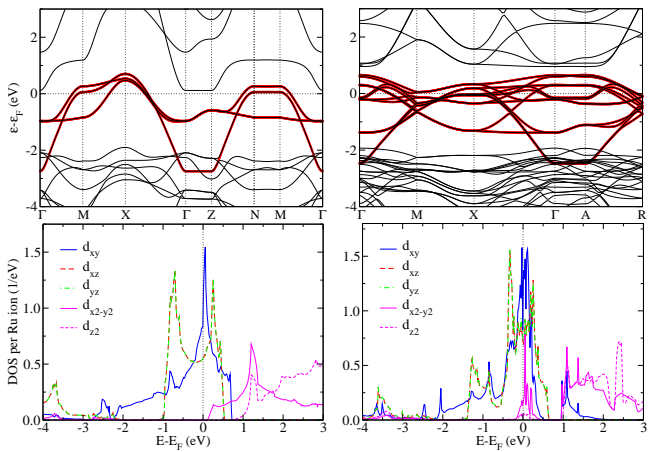

\centering
\includegraphics*[width=4.25cm]{bands-214.eps}\hspace*{-0.1cm}
\includegraphics*[width=4.25cm]{bands-327.eps}\\
\includegraphics*[width=4.25cm]{dos-214.eps}\hspace*{-0.1cm}
\includegraphics*[width=4.25cm]{dos-327.eps}
\caption{(Color online) Comparison of band structure (top) and local Ru($4d$) 
DOS (bottom) on the LDA level between Sr$_2$RuO$_4$ (left) and Sr$_3$Ru$_2$O$_7$ 
(right) when neglecting SOC. 
The red curves in the top pictures display the Wannier-like dispersion
as obtained from the MLWF scheme. The angular-momentum resolution is obtained from 
projecting the Bloch states onto cubic harmonics with a radial extension of 2.0 a.u..
\label{fig:bandos}}
\end{figure}
Lets start by picturing the interacting electronic systems within Sr$_2$RuO$_4$ 
and Sr$_3$Ru$_2$O$_7$ at equilibrium for ${\bf H}$=0. Besides the different number
of Ru ions in the primitive unit cell, there are important differences in the 
crystal symmetry (see Fig.~\ref{fig:214-327}). The $n$=1
compound has ideal tetragonal symmetry (space group $I4/mmm$) with the 
fourfold rotation $C_4^z$ around the $c$-axis. On the contrary, the $n$=2 ruthenate 
shows orthorhombic symmetry (space group $Bbcb$), whereby the RuO$_6$ octahedra 
display a small rotation of $6.8^{\circ}$ in the $ab$-plane~\cite{sha00}, resulting 
only in the twofold symmetry element $C_2^z$. Basic structural and electronic 
differences may also be understood from an $\sqrt{2}$$\times$$\sqrt{2}$ 
reconstruction of the Sr$_2$RuO$_4$ unit cell within the $ab$-plane.
\begin{figure}[t]
\centering
\includegraphics*[width=8.5cm]{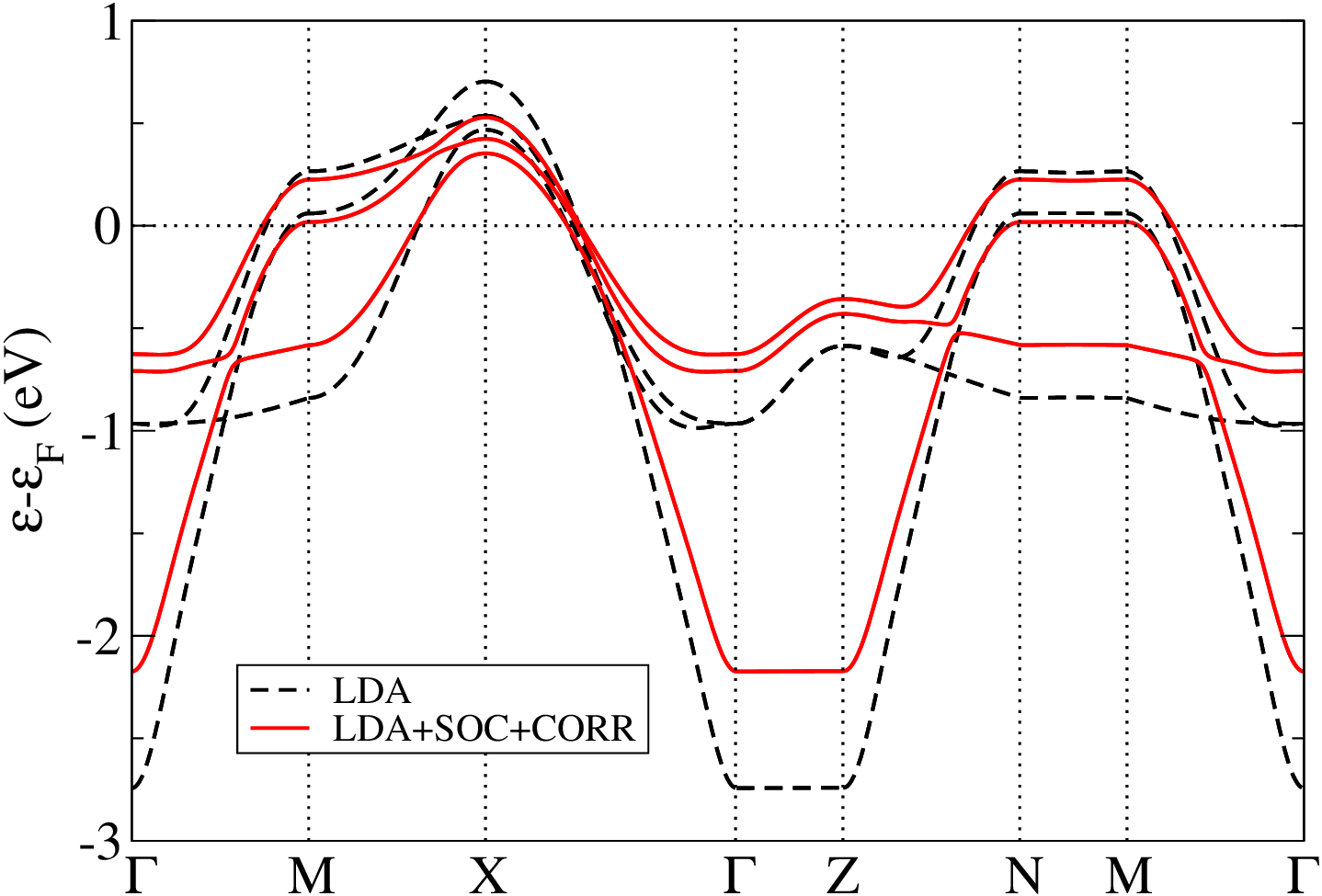}
\caption{(Color online) Sr$_2$RuO$_4$ quasiparticle band structure within 
standard LDA as well as with including SOC and correlations.
\label{fig:214lda_soc_corr}}
\includegraphics*[width=8.5cm]{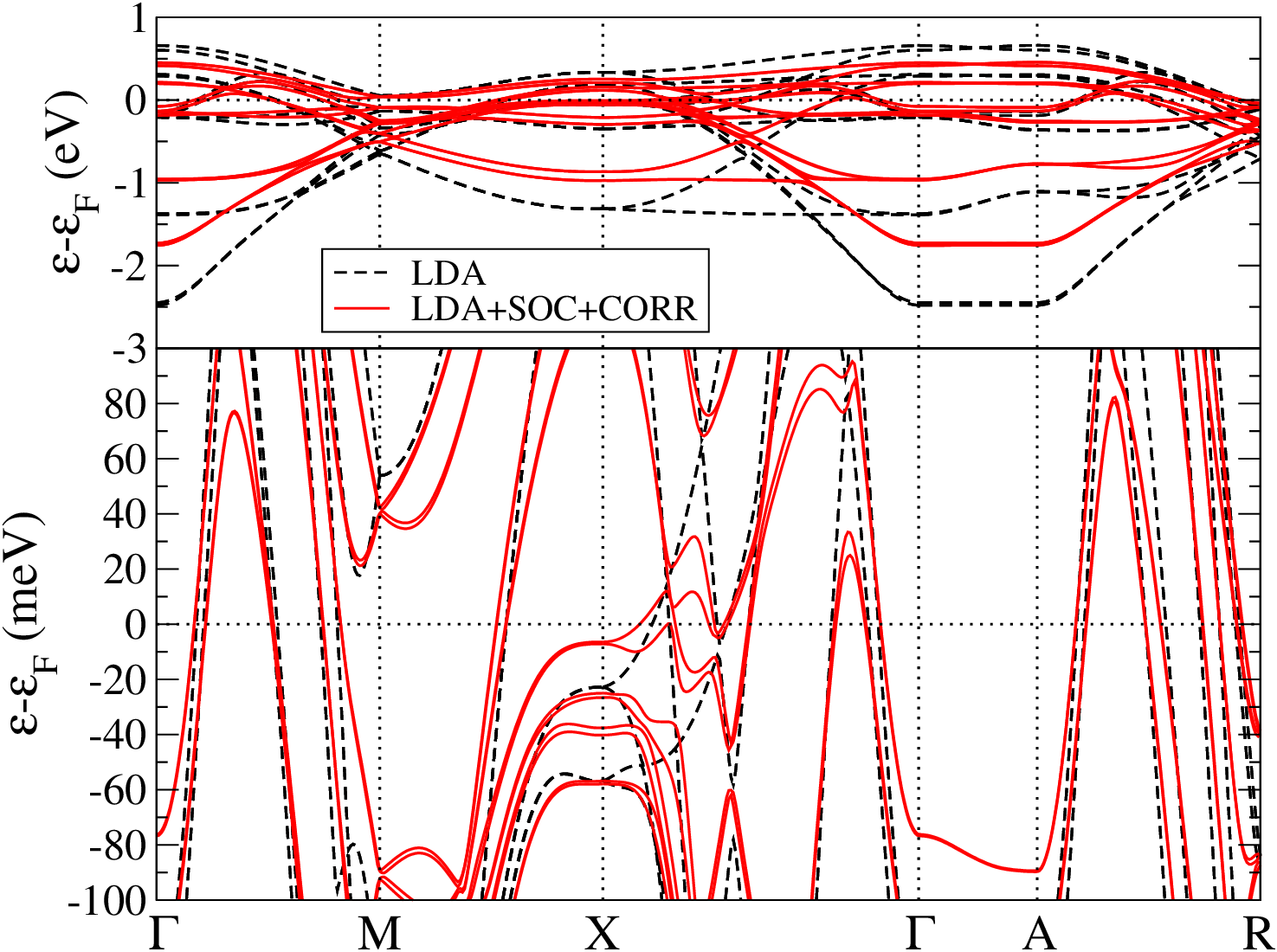}
\caption{(Color online) Same as Fig.~\ref{fig:214lda_soc_corr}, here for
Sr$_3$Ru$_2$O$_7$. The lower panel provides a blow up of the dispersions close to the
Fermi level.
\label{fig:327lda_soc_corr}}
\end{figure}
\begin{table}[h]
\begin{tabular}[t]{C{2cm}|C{2cm}|C{2cm}|C{2cm}}
    & $J_{\rm H}$=0.20 &  $J_{\rm H}$=0.35 & $J_{\rm H}$=0.50\\ \hline
    $U$=1.5 & -13.9 & -7.8 & -4.4\\
    $U$=2.0 & -13.6 & -6.5 & -2.8\\
    $U$=2.5 & -11.1 & -3.8 & -0.6\\
\end{tabular}
\caption{Energy (in meV) of the highest occupied band at the $X$ point in Sr$_3$Ru$_2$O$_7$ 
for different $U$ and $J_{\rm H}$ combinations (both in eV). In LDA without spin-orbit 
coupling that energy amounts to -22.8 meV.}
\label{tab:pardep}
\end{table}

Figure~\ref{fig:bandos} depicts the band structure and the local Ru($4d$) density of
states (DOS) of the two layered compounds as retrieved from conventional LDA 
calculations without SOC. The low-energy regime is dominated by the $t_{2g}$ manifold 
of the Ru$(4d)$ shell and can be downfolded to associated Wannier-like states. In
both systems the latter leak into the oxygen-dominated block of bands. The overall 
$t_{2g}$ bandwidth of the bilayer compound is slightly smaller ($\sim 3.1$ eV) than 
for the single-layer system ($\sim 3.4$ eV). The $e_g$ contribution close to the 
Fermi level is minor for Sr$_2$RuO$_4$, however the $d_{x^2-y^2}$ character is 
non-negligible close to $\varepsilon_{\rm F}$ for Sr$_3$Ru$_2$O$_7$,~\cite{tam08} 
as may be seen from the local DOS. Note that this contribution is included in the 
effective $t_{2g}$-MLWF construction.~\cite{pie11} In this respect it is important 
to record that we use in the following the ($xy$, $xz$, $yz$) terminology, though the 
corresponding orbitals are only $t_{2g}$-like in the sense of the present minimal 
MLWF construction. Note that these orbital functions (as true low-energy
states) are pointing inbetween the Ru-O-Ru bonds of the in-plane square lattice
(compare also Fig.~\ref{fig:direct}).

In the single-layer compound the energetics of the $t_{2g}$ manifold is split into 
($d_{xz}$, $d_{yz}$) and $d_{xy}$. The $d_{xz}$, $d_{yz}$ orbitals are truly 
degenerate, with crystal-field splitting $\Delta_{xy}$=$\varepsilon_{xz,yz}$$-$$\varepsilon_{xy}$=113 meV to the $d_{xy}$ Wannier level. With very small quantitative 
differences, Sr$_3$Ru$_2$O$_7$ has quasi-degenerate $d_{xz}$, $d_{yz}$ levels 
($\Delta$=0.3 meV) and $\Delta_{xy}$=115 meV. Hence it has to be emphasized that
there is already a small but nonzero splitting between $d_{xz}$ and $d_{yz}$, giving
rise to nominally slightly different local occupations. Note also that the strong 
local-DOS differentiation between ($xz$, $yz$) and $xy$ does no longer hold in the 
bilayer compound. Albeit the $d_{xy}$ Wannier level is always lower in energy, the 
local orbital electron fillings in LDA for $(d_{xy},d_{xz},d_{yz})$ are 
(1.24, 1.38, 1.38) in the case of $n$=1 and (1.40, 1.30, 1.30) for $n$=2. Thus there 
is a change in the $t_{2g}$ occupation hierachy between both ruthenates due to 
band-dispersion effects. Right at the Fermi level, the DOS of Sr$_2$RuO$_4$ is 
close to a van-Hove singularity slightly above $\varepsilon_{\rm F}$, while in the 
case of the bilayer the Fermi level is located in a large dip of a complicated 
multi-valley-peak DOS at low-energy. 
\begin{figure}[t]
\centering
\includegraphics*[width=7.5cm]{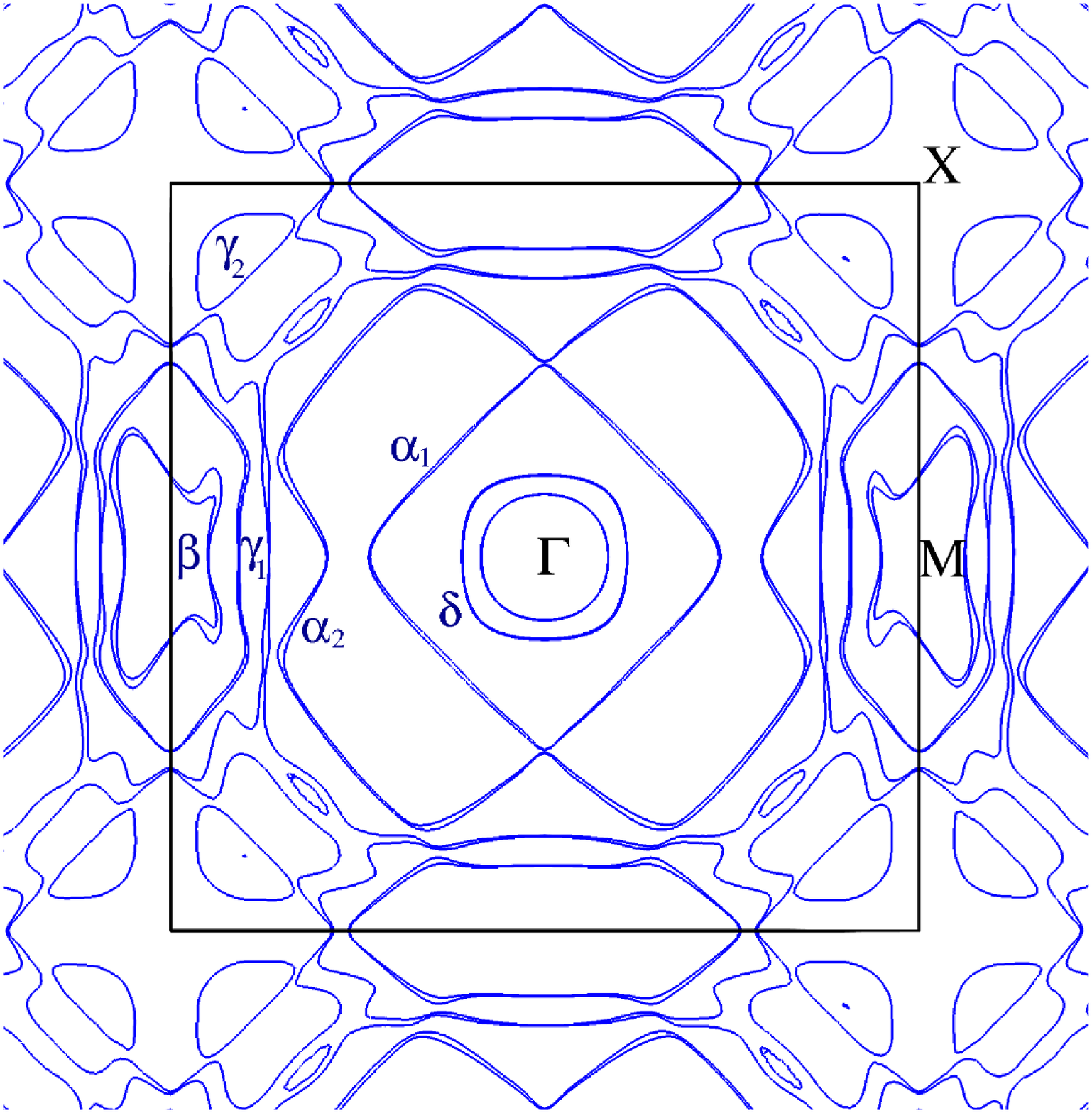}
\caption{(Color online) Interacting Fermi surface for Sr$_3$Ru$_2$O$_7$ in the
basal $k_z$=0 plane from extended-LDA including SOC and electronic correlations. 
The black square marks the BZ cut and the labeling of the different sheets is 
according to Ref.~\onlinecite{tam08}.\label{fig:fs327}}
\end{figure}

Solving the problem posed by the minimal Hamiltonian~(\ref{eq:fullham}) for each 
system results in modifications in the low-energy dispersions. 
Figure~\ref{fig:214lda_soc_corr} shows the $t_{2g}$-like QP band structure 
of Sr$_2$RuO$_4$ from the extended-LDA treatment. One observes the expected combined
main features already known from the existing separate SOC~\cite{pav06,hav08,roz11} 
and correlated~\cite{lie00,pch07,mra11,mal11} studies. Namely the lifting of 
degeneracies, e.g. close to the $\Gamma$ point for the bands with dominant 
$xz,yz$ character, also resulting in now avoided band crossings, e.g. close to the 
$X$ point, when including SOC. Electronic correlations lead in the present 
approximation to band-narrowing and -shifting. Lifetime effects as well as incoherent 
spectral weight can not be retrieved within RISB at saddle-point. But the method
captures very well the slightly modified Fermi-level band crossings in e.g. the 
$\Gamma M$ direction as well as the shift of the van-Hove singularity towards the
$M$ point.~\cite{dam00,she01,lie00}
The band renormalizations are substantial, however not quite as strong as obtained 
within DMFT calculations with more elaborate frequency dependence of the 
self-energy~\cite{lie00,pch07,mra11} to match the ARPES measurements. In the present
modeling quantum fluctuations are missing and to obtain a ratio $m^*/m_{\rm LDA}$ of 
the order of 3-4 for Sr$_2$RuO$_4$, in good comparison with photoemission~\cite{she01} and 
dHvA~\cite{mac96} data, a value $U$$>$3 eV would be needed.
\begin{figure}[b]
\centering
\includegraphics*[width=8.5cm]{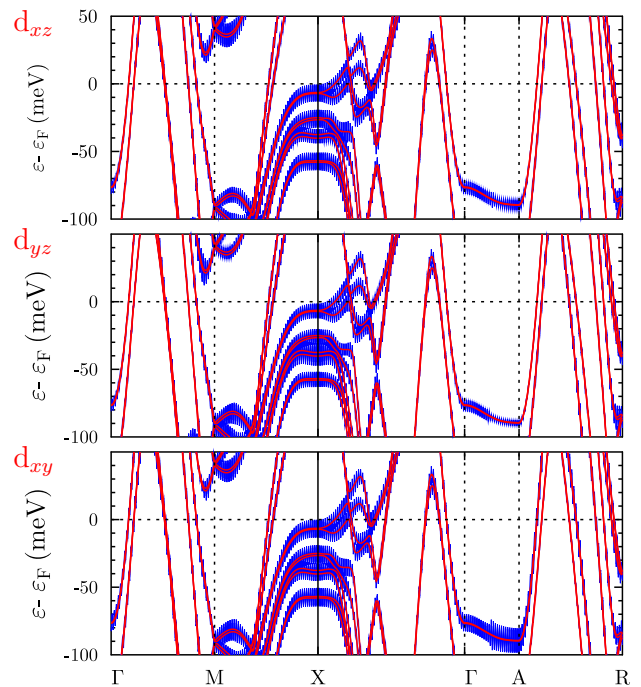}
\caption{(Color online) Individual weights of the $t_{2g}$ Wannier orbitals on
the low-energy QP bands from the extended-LDA calculation. 
\label{fig:fat}}
\end{figure}

Accordingly, Fig.~\ref{fig:327lda_soc_corr} depicts the comparison between the 
quasiparticle bands (including SOC and correlations) with the conventional LDA 
dispersions for the bilayer Sr$_3$Ru$_2$O$_7$. Due to the reduced symmetry the level
of complexity is now surely raised. The complicated low-energy manifold with its 
rather flat bands already on the LDA level now shows significant effective-mass 
renormalization and additional splittings within the extended electronic structure
examination. From experiment,~\cite{tam08,mer10} the ratio $m^*/m_{\rm LDA}$ is 
effectively on the order of 6, thus even larger than in the single-layer compound.
Most interestingly, the blow up of the computed Fermi-level neigborhood renders
it obvious that especially close to the $X$ point severe changes take place. For 
instance a two-fold band approaching from $M$ and splitting when passing $X$ is
strongly shifted to a low-energy of about 7 meV in the occupied part. 
Importantly, the latter small scale may only be reached by the combination of SOC 
{\sl and} explicit Coulomb interactions. While the spin-orbit coupling is 
responsible for the appearance of split-up bands, the electronic correlations provide 
the narrowing and further shifting towards $\varepsilon_{\rm F}$. Increasing $U$ 
and $J_{\rm H}$ leads to an even closer placement nearby the Fermi level (see 
Tab.~\ref{tab:pardep}). Hence the present approach is truly capable of describing the
very low-energy scale the bilayer compound is famously known for. Furthermore the key 
features of the named dispersive structure around $X$ with its local minima and maxima 
are in close agreement with results from ARPES studies by Tamai {\sl et al.}.~\cite{tam08} 
The Fermi-level crossings in this region of the Brillouin zone (BZ) give rise to the 
so-called $\gamma_2$ pocket, which is under strong suspicion to play a vital role in 
the peculiar metamagnetic behavior of this compound.~\cite{pue10} The complete Fermi 
surface (FS) obtained from our extended-LDA calculations exhibited in 
Fig.~\ref{fig:fs327} is in good accordance with the one determined from 
photoemission.~\cite{tam08} Although however from our study the $C_2^z$-symmetry 
character seems vital in the fermiology, but note that the experimental data in 
Ref.~\onlinecite{tam08} is symmetrized along the $\Gamma X$ direction.
\begin{figure}[t]
\centering
\includegraphics*[width=8.5cm]{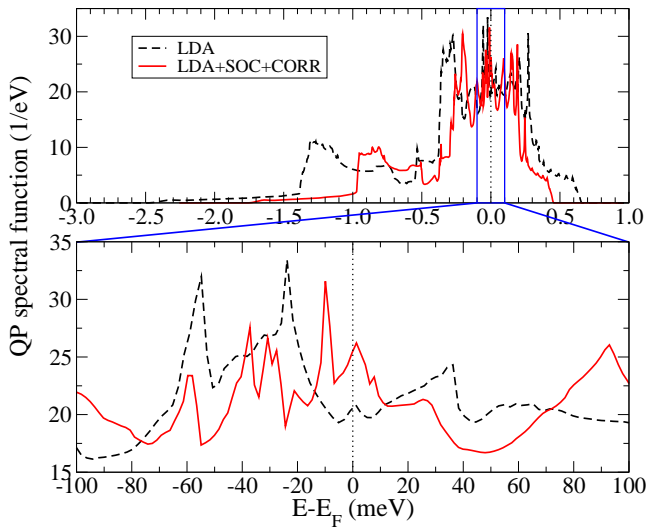}
\caption{(Color online) Interacting quasiparticle DOS for Sr$_3$Ru$_2$O$_7$ compared
to the LDA result. The QP content including SOC and correlations is scaled with
$Z$=0.7 to account for the modified integrated spectral weight resulting  
from the slave-boson framework. The bottom panel shows a blow up around the Fermi 
energy.\label{fig:327dos}}
\end{figure}

An important question concerns the band characters at low energy in order to touch
base with the local-orbital viewpoint. Figure~\ref{fig:fat} therefore shows the
so-called fatbands (orbital weight proportional to an artifical band broadening) for
the $t_{2g}$ manifold. Its easily seen that the bilayer system is far from being a 
textbook example when it comes to attributing bands to a certain azimuthal quantum
numbers, since e.g. the supposingly relevant bands close to $X$ are of strong mixed 
$t_{2g}$ character. Nonetheless small asymmetries may be identified. The topmost
occupied band at $X$ has somewhat more $d_{xz}$ than $d_{yz}$ weight, true also
for the lowest one in the given energy window. In between there is only one with more
$d_{yz}$ character. The $d_{xy}$ orbital has notably overall the same order of 
weight in this region of the BZ as its out-of-plane companions. Interestingly,
from the strong weight along $\Gamma A$ the $d_{xy}$ character seems to dominate the 
propagation perpendicular to the RuO$_6$ planes.
\begin{figure}[t]
\centering
 \includegraphics*[width=8.5cm]{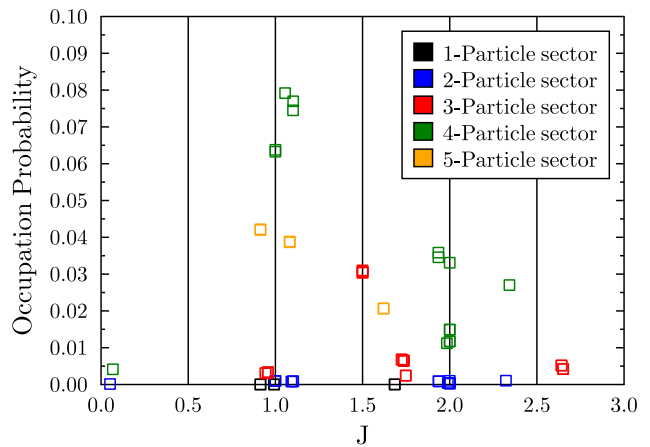}
\caption{(Color online) Histogram showing the occupation probabilities of calculated 
multiplets in Sr$_3$Ru$_2$O$_7$ ordered by the total angular momentum $J$.}
\label{fig:histjmult}
\end{figure}

The low-energy scales may also be confirmed from the quasiparticle DOS plotted in
Fig.~\ref{fig:327dos}. Therefrom its again obvious that the states close to
$\varepsilon_{\rm F}$ are strongly pronounced in extended-LDA, shifting prominently
to the Fermi energy. The LDA-DOS exhibits a smaller peak at the Fermi level 
within a valley of $\sim$60 meV width. With the additional interactions that feature
is strengthened and importantly a peak in the low-energy occupied region is sharpened
and shifted towards $\varepsilon_{\rm F}$, being located at $\sim$10 meV. These 
findings of increased spectral weight below $\varepsilon_{\rm F}$ within a meV range 
is in accordance with photoemission studies~\cite{tam08} and also supported from
specific-heat data.~\cite{ros10} 

Finally, Fig.~\ref{fig:histjmult} depicts the occupation probabilities of the local
$t_{2g}$-based multiplets according to the converged slave-boson amplitudes of the lattice
calculation in the metallic state. There are various sizeable multiplet weights, 
non-surprisingly with an overall domination of the ones from the four-particle sector. 
The atomic ground-state multiplet with $L$=$S$=$J$=1, an orbital and spin triplet, has
also the largest weight in the itinerant regime. Note that the deviations of $J$ 
from the ideal values results from the small $e_g$ weight inter-mixing within
the LDA-derived Kohn-Sham Hamiltonian based on the low-energy $t_{2g}$-like orbitals.
For Sr$_2$RuO$_4$ the corresponding picture looks very similar, also there the 
$L$=$S$=$J$=1 multiplet has the maximum weight.

\section{Bilayer ruthenate in applied magnetic field}
\begin{figure}[b]
\centering
\includegraphics*[width=8cm]{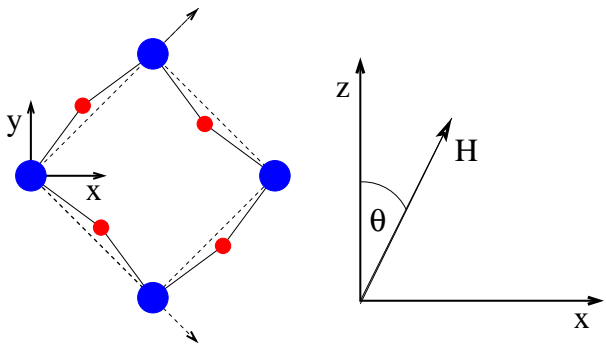}
\caption{(Color online) Left: view along the $c$ axis of sketched Sr$_3$Ru$_2$O$_7$ 
in order to clarify the invoked cartesian coordinate system with Ru ions in 
(blue/dark) and O ions in (red/grey). Right: applied magnetic-field direction where 
the $z$ axis equals the crystallographic $c$ axis.\label{fig:direct}}
\end{figure}

Since our approach is in the position to correctly address the equilibrium low-energy 
correlated electronic structure, we expect a qualitatively meaningful description of 
Sr$_3$Ru$_2$O$_7$ with applied magnetic field ${\bf H}$. The Zeeman-type
local interaction is now included together with the spin-orbit and Coloumb 
interactions. All are adequately treated due to the generality of our formalism, 
also allowing for arbitrary field directions. Note that the crystal structure is 
constructed such that 
the in-plane square lattice evolves along the Ru-O-Ru bonds, while the original 
$x$-, $y$-axes point inbetween these bonding directions (see Fig.~\ref{fig:direct}). 
In the present investigation the field direction is modified within the $xz$ plane, 
i.e., the tilting of $H$ takes place along the diagonal of the square lattice which 
also agrees with the orthorhombic $a$ axis. Notably the $x$ direction in real space 
corresponds to the $\Gamma M'$ ($M$ rotated by 90$^{\circ}$ degrees) direction in 
reciprocal space and therewith the $\Gamma X(X')$ direction amounts to QP propagation 
along the Ru-O-Ru bond on the rods of the square lattice. 
The resulting net magnetic moment $M$ per Ru ion is composed of spin and 
orbital-momentum parts, i.e. $M$=$\mub\langle g\,\bJ$$\cdot$$\bH\rangle$ and is here 
computed on the local level from the self-consistent slave-boson amplitudes 
(see eq.~(\ref{eq:locex})) and not from $k$-integrating the associated QP 
contributions. While in effective single-particle calculations both approaches would
yield identical results, in slave-boson theory those numbers may in principle differ
since the QP part only carries the contribution to the itinerant character of the
electron. Via eq.~(\ref{eq:locex}) it is ensured that the occupation number of the
physical electron is retrieved.

Because of the low-energy scales involved in this problem, especially
for the close-to-realistic investigation of the magnetic behavior the numerics
is however in any case far from simple. The complicated local interacting Hamiltonian 
asks for about 2000 slave-boson amplitudes to solve for, interlinked with a 
24$\times$24 QP-Hamiltonian problem (four Ru ions with three orbitals per ion, allowing for
spin degrees of freedom within each orbital) on a properly dense $k$-point mesh.

\subsection{The case ${\bf H}\parallel {\bf c}$}
\begin{figure}[h!]
\includegraphics*[width=8.5cm]{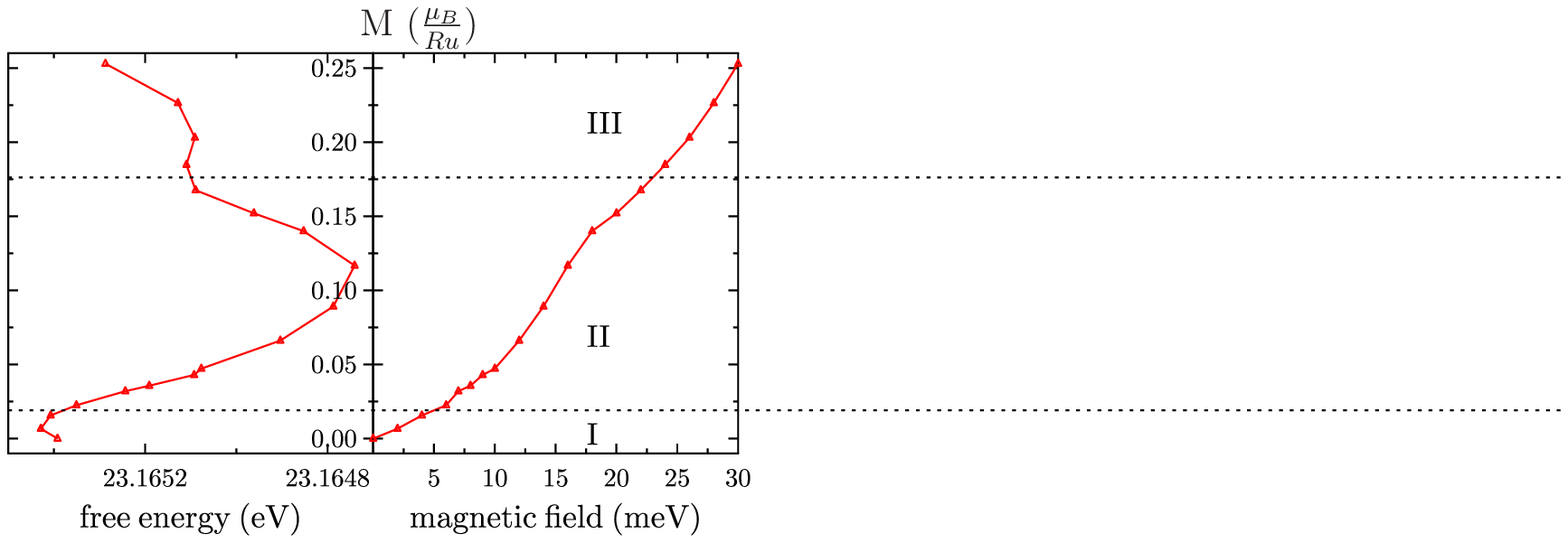}
\caption{(Color online) Sr$_3$Ru$_2$O$_7$ in applied magnetic field (measured in
meV) along ${\bf z}$. Left: free energy vs. net magnetic moment, 
right: net magnetic moment per Ru ion.\label{fig:mmfreez}}
\includegraphics*[width=8cm]{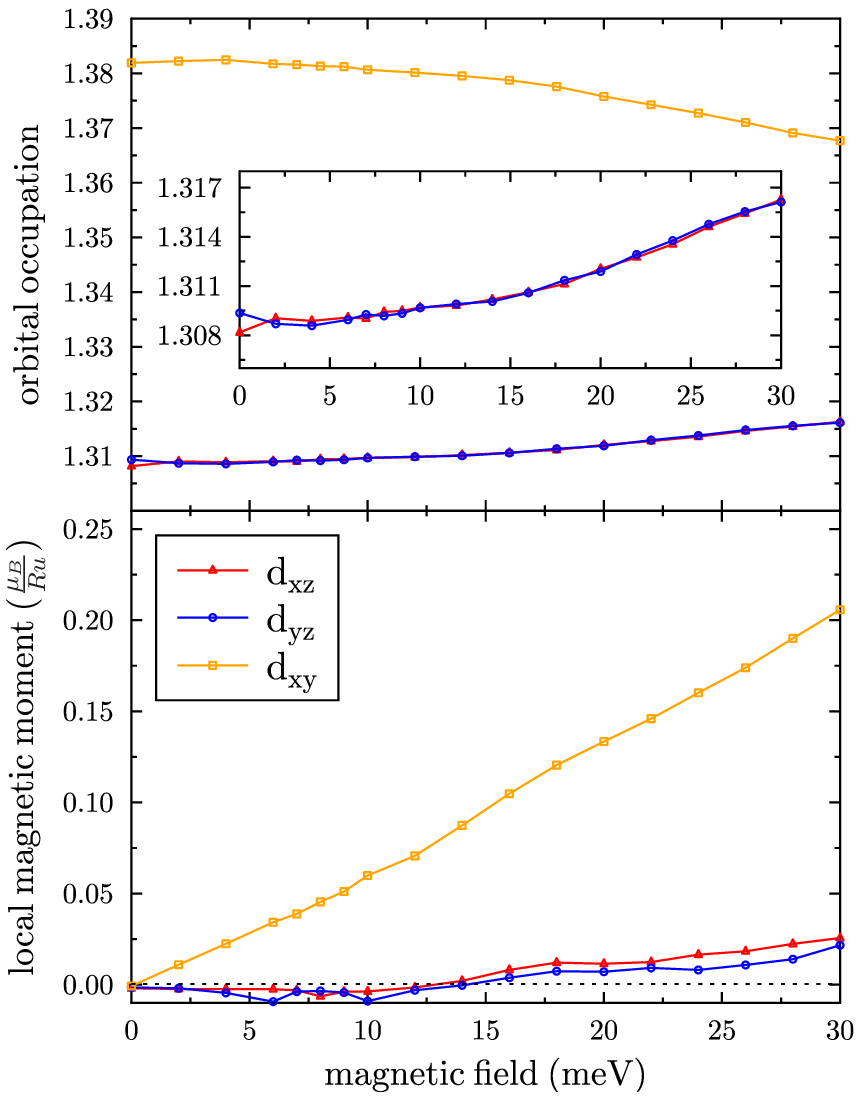}
\caption{(Color online) Orbital-resolved  local occupation (top) and contributions 
to the Ru net magnetic moment (bottom) with magnetic field. The inset shows a blow-up
of the $d_{xz}$, $d_{yz}$ curves.
\label{fig:orbdep}}
\end{figure}
We first choose the crystallographic $c$-axis (i.e., the $z$ direction) perpendicular 
to the RuO$_2$ planes for the magnetic-field direction, i.e. ${\bf H}$=$H\,\hat{\bf c}$. 
Figure~\ref{fig:mmfreez} shows the evolution of $M$ with increasing field
strength together with the obtained total free energy for the bilayer system. Both
curves display a rather non-trivial behavior. In the following the values of $H$ are 
given in meV. Up to $H$=5 the value of $M$ rises linearly (region I), with however 
negative free-energy curvature, hinting towards unfavorable field penetration. In the 
range 5$\lesssim$$H$$\lesssim$23 a stronger rise of $M$ occurs, followed by a
non-monotonic behavior (region II). In that second region the free energy is first
depleting and then again rising along with the non-monotonic part, overall evolving
with the positive curvature of a stable phase. Lets also note that it appears as if 
in the first part 5$\lesssim$$H$$\lesssim$10 the $M$ evolution as well as the 
free-energy curve display some non-trivial modulation. Finally for even larger $H$ 
(region III) the magnetic moment enhances further with close-to-linear development.
In III the free energy follows a novel parabolic shape at higher values than in II.
The described evolution reveals the metamagnetic behavior of $M$, with strong 
resemblance to the experimental data~\cite{perr01,ros09}. Notably, the observed 
metamagnetism is again only obtained in the present computations if both, SOC 
{\sl and} electronic correlations, are included. Due to the strong signatures in the 
free energy, the transition between regions I/II as well as regions II/III are 
clearly of first order. We however did not investigate the transition orders in
a more elaborate fashion (e.g. via computing the Hesse matrix).

To connect these global results to the orbital degrees of freedom, 
Fig.~\ref{fig:orbdep} depicts the orbital-resolved local occupations and 
contributions to the magnetic moment. With increasing field the 
$d_{xy}$ filling shrinks, while the one for $d_{xz}$, $d_{yz}$ grows. Thus notably
there is an inter-orbital charge transfer from $d_{xy}$ to $d_{xz}$, $d_{yz}$ with
magnetic field. Whereas for $H$=0 a marginal filling difference between the
latter orbitals is observed, with growing magnetic field the occupations of the
quasi-degenerate levels more or less align. The respective orbital contributions to 
the magnetic moment within the $t_{2g}$ manifold are strongly varying. While the 
dominant $d_{xy}$ part shows substantial paramagnetic response with field, the 
generally much smaller $d_{xz}$, $d_{yz}$ terms exhibit intricate behavior.
 They start with 
flat, nearly constant minor diamagnetic response for small magnetic field and both 
only turn into weak paramagnetic characteristics at $H$$\sim$13. Interestingly,
this PM behavior shows small differences in the amplitude for $d_{xz}$ and $d_{yz}$
(with $M_{xz}$$>$$M_{yz}$), with even an observable sudden increase in that 
difference when the II/III transition occurs. As the sole $d_{xy}$ response shows no 
definite MM signals, the para/dia discrimination in the orbital response of $d_{xz}$, 
$d_{yz}$ appears as a key microscopic building block for the MM behavior.
\begin{figure*}[t]
\onecolumngrid
\centering
\includegraphics*[height=11.5cm]{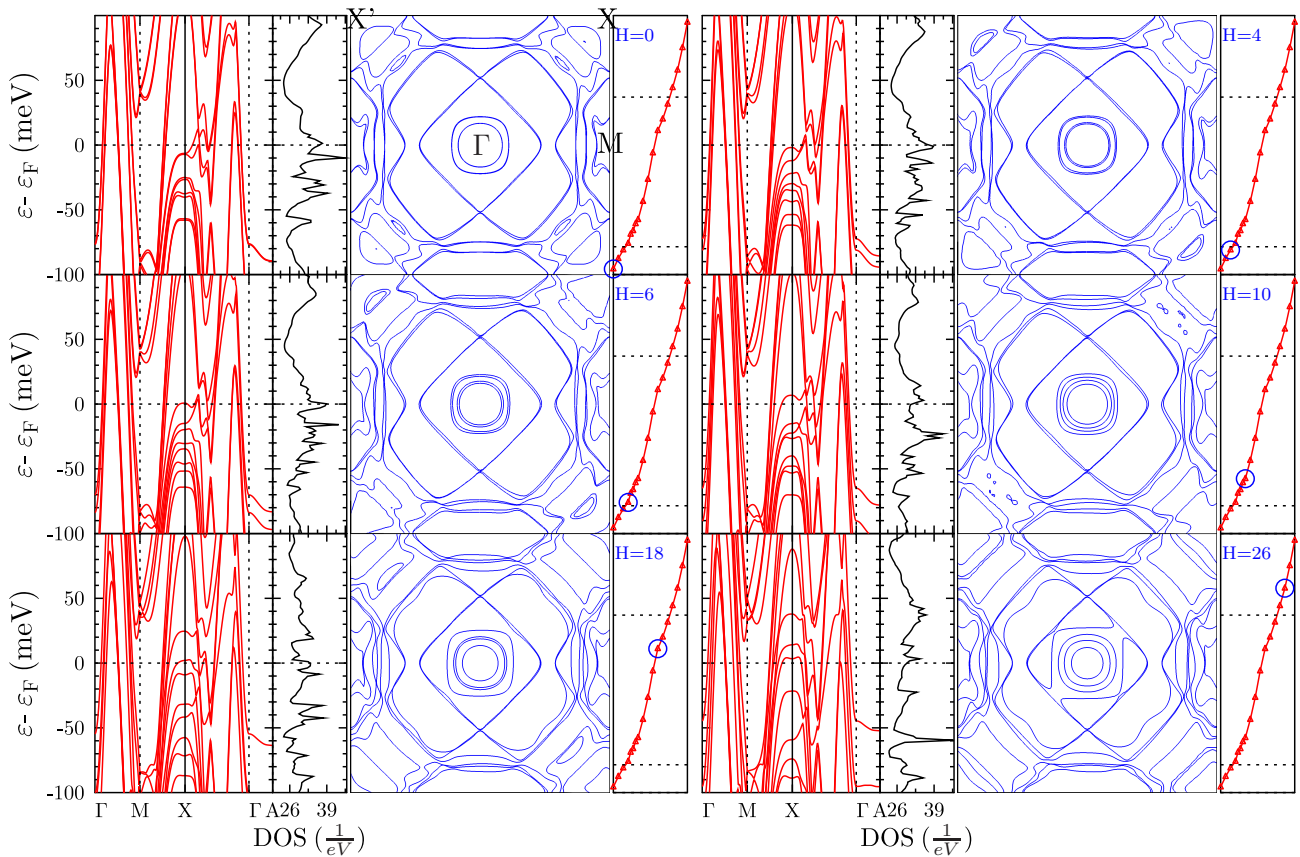}
\caption{(Color online) Development of the QP band structure, DOS and
Fermi surface with magnetic field along the $z$ direction. The (blue/grey) circle on 
the right panel always marks the respective $H$ value.
\label{fig:bands-field-z}}
\twocolumngrid
\end{figure*}

The orbital discrimination in the magnetic behavior already provides a first clue to
the MM puzzle of the bilayer system. A second important insight originates from 
following the development of the QP bands with magnetic field, presented in 
Fig.~\ref{fig:bands-field-z}. It is seen that increasing $H$ amounts to intriguing 
changes in the low-energy manifold, most notably to relevant shifts in the peculiar 
QP structure close the $X$ point. For $H$=4, non-surprisingly there are further band 
splittings along $X$ compared to the case of zero magnetic field and the lowest-energy
$p$-shaped band at this $k$-point is now placed just below $\varepsilon_{\rm F}$. 
A discrimination between the $\gamma_2$ pockets around $X$ and $X'$ is furthermore
clearly visible. Going to $H$=6 the named band is locked to the Fermi level and 
$\gamma_2$ at $X$ has opened towards the BZ boundary where as $\gamma_2$ at $X'$ has 
shrunk. 
For $H$=10 the former band is above $\varepsilon_{\rm F}$ and both $\gamma_2$ 
pockets are opened. At $H$=18 the second-lowest band at $X$, $X'$ crosses the Fermi level 
and a minor pocket structure reappears at $X'$. Morover the $\alpha_2$ sheet starts 
to become increasingly distorted along $\Gamma X(X')$ and also begins effectively 
shrinking with growing field. The latter signatures are strengthened for $H$=26 with 
the additional onset of hybridization between $\alpha_1$ and $\delta$. 
The pockets close to $X,X'$ remain both opened in that large-field region III, with
now three low-energy bands having crossed $\varepsilon_{\rm F}$ at $X$,$X'$. Along with 
these changes, the QP-DOS of course runs through several peaks at the Fermi energy, but
evidently exhibiting an evolution different from a pure shifting of the $H$=0 
structure. Thus from the Bloch perspective the picture of Lifshitz transitions 
underlying the magnetic response emerges. Various authors have already pointed out 
the importance of van-Hove singularities crossing the Fermi level and we here can 
verify this mechansim based on the complete realistic starting point. The fact that 
not only the $\gamma_2$ sheet but also the inner sheets, most notably $\alpha_2$, 
may play a vital role in the MM response was also retrieved in a recent experimental 
study employing spectroscopic imaging scanning tunnelling microscopy.~\cite{lee09}

\begin{figure}[t]
\centering
\includegraphics*[width=8.25cm]{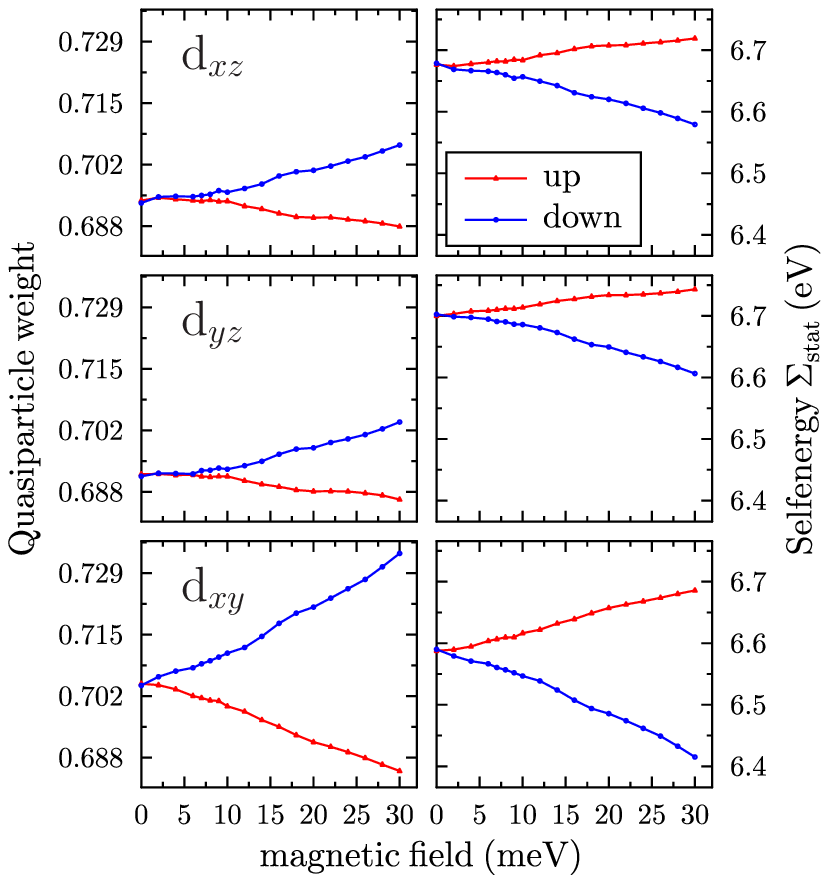}
\caption{(Color online) Orbital-dependent QP weight (right) and static 
self-energy (left) with $H$.
\label{fig:self-z}}
\includegraphics*[width=8.25cm]{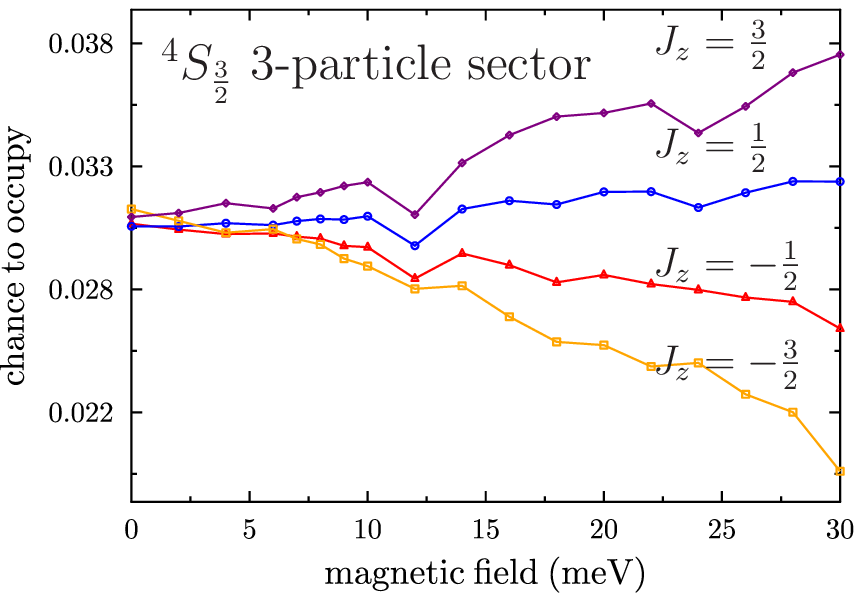}
\caption{(Color online) Splitting of the $^4S_{\nicefrac{3}{2}}$ multiplet 
in the applied field.\label{fig:multiplet-z}}
\end{figure}

The applied magnetic field leaves also some signatures in the orbital-dependent
electronic self-energy (see Fig.~\ref{fig:self-z}). All orbital sectors display the 
expected splitting in the QP weight $Z$ and static self-energy $\Sigma^{\rm stat}$
due to the spin-filling inbalance with larger magnetic polarization. The splitting
for $d_{xz}$, $d_{yz}$ is weaker and especially minor at small field where the 
nearly constant diamagnetic response occurs. Overall there is no strong modification
of the correlation strength with $H$, the MM signatures show up somewhat stronger
in the QP weight that is associated with band renormalizations. From the calculation, 
the value for $Z_{xy}$ is slightly larger than for the remaining two $t_{2g}$ 
orbitals. But because of the intriguing hybridizations (compare Fig.~\ref{fig:fat}) no 
trivial relation may be drawn therefrom in view of the respective renormalized 
effective masses on the various Fermi sheets. Our RISB formalism is furthermore able 
to reveal the local-multiplet behavior of the correlated system. 
Via the slave-boson amplitudes the method allows to evaluate the occupation 
probability of a given eigenstate of the local Hamiltonian (\ref{eq:hamloc}) within 
the complete itinerant solution. 
For instance, Fig.~\ref{fig:multiplet-z} depicts 
the splitting characteristics of the $^4S_{\nicefrac{3}{2}}$ multiplet from the 
three-particle sector assoicated with orbital momentum $L$=0, i.e. having ${\bf J}$ 
as purely spin-defined. As exptected, the $J_z$ degeneracy is lifted for $H$$\not=$0, 
showing non-trivial signature close to the phase transitions between the 
different region I-III. Sure enough, the four-particle sector is most dominantly
occupied for the Ru($4d$) $t_{2g}$ shell, but the multiplets there do not exhibit a very 
conclusive behavior with applied $H$.

\subsection{The case ${\bf H}\nparallel {\bf c}$} 
\begin{figure}[b]
\centering
\includegraphics*[width=8.5cm]{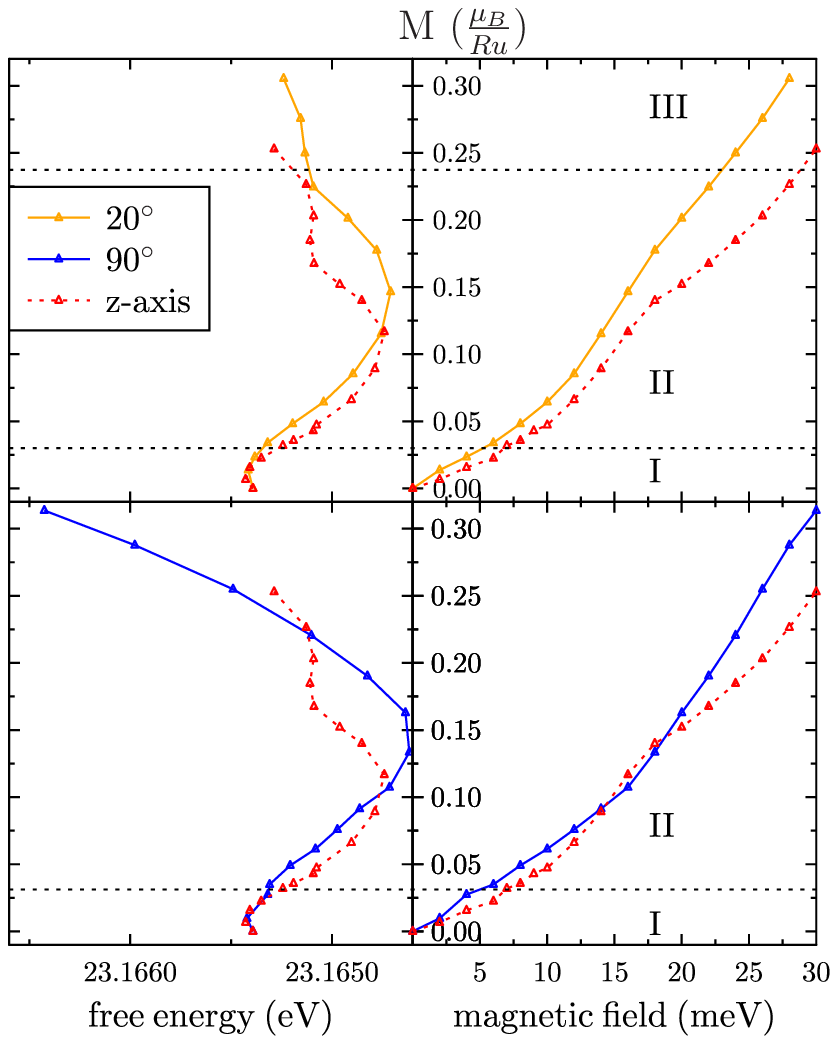}
\caption{(Color online) Free energies (left) and net magnetic moment (right) 
for $\theta$=20$^\circ$ (top) and 
$\theta$=90$^\circ$ (bottom), compared to the moment for $H$ along the $z$ direction
($\theta$=0$^\circ$), respectively
\label{fig:mom20-z}}
\includegraphics*[width=8.5cm]{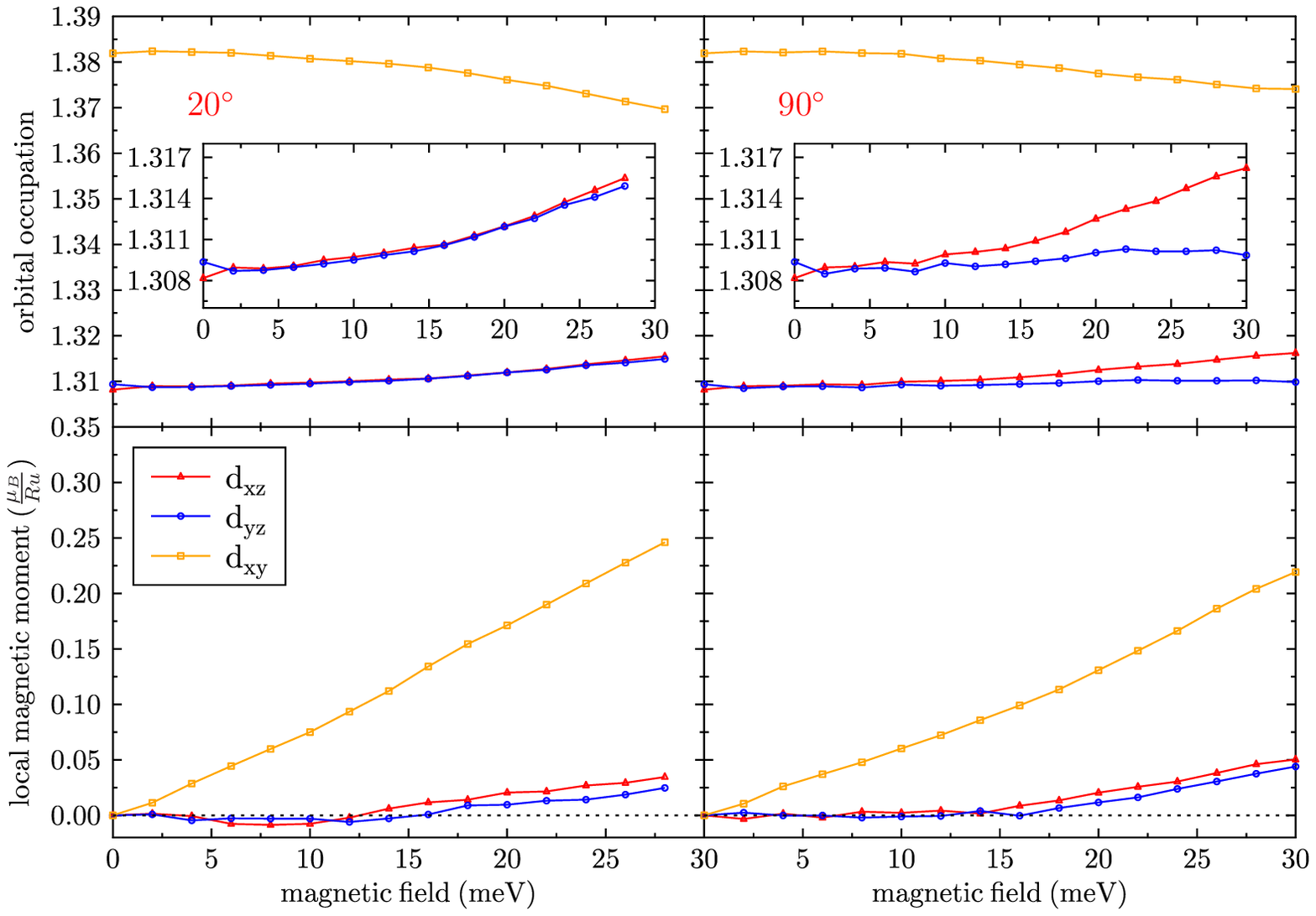}
\caption{(Color online) Orbital-resolved occupation and contributions to
the magnetic moment as in Fig.~\ref{fig:orbdep}, here for $\theta$=20$^\circ$ (left) 
and $\theta$=90$^\circ$ (right).\label{fig:orbdep20x}}
\end{figure}

In principle our approach works for arbitrary polar angles $\theta$ between {\bf H} 
and the $c$-axis of the system. However since the computations are rather expensive 
we have chosen here only two additional specific values (besides $\theta$=0), namely 
$\theta$=20$^{\circ}$ and $\theta$=90$^{\circ}$. The azimuthal angle is put to zero, 
i.e. the latter in-plane magnetic field points along the $x$ direction inbetween
the Ru-O-Ru bond of the square lattice (see Fig.~\ref{fig:direct}). 
The evolution of the net magnetic moment per Ru ion for the two new field angles 
together with the respective free-energy plot is displayed in Fig.~\ref{fig:mom20-z}.
When directly comparing the magnetizations for $\theta$=20$^{\circ}$ with the former
$\theta$=0, one first realizes that for a given $H$ the value of $M$ is increased.
This seems to be in line with the experimental data from the work of Grigera 
{\sl et al.}~\cite{gri03} showing an enhancement of the real part of the differential
susceptibility at the MM transition with $\theta$ (note that $\theta$ is defined as 
the angle between field and the $ab$-plane in Ref.~\onlinecite{gri03}).
The overall phenomenology of $M(H)$ for $\theta$=20$^{\circ}$ is still rather 
similar to the case ${\bf H}$$\parallel$${\bf c}$. Note that although the upper 
first-order transition happens at larger magnetic moment for $\theta$=20$^{\circ}$, 
the value corresponds to nearly the same magnetic-field strength $H$. On the contrary,
for ${\bf H}$$\perp$${\bf c}$ along $x$ the overall characteristic is qualitatively 
different.
\begin{figure*}[t]
\onecolumngrid
\centering
\includegraphics*[height=15cm]{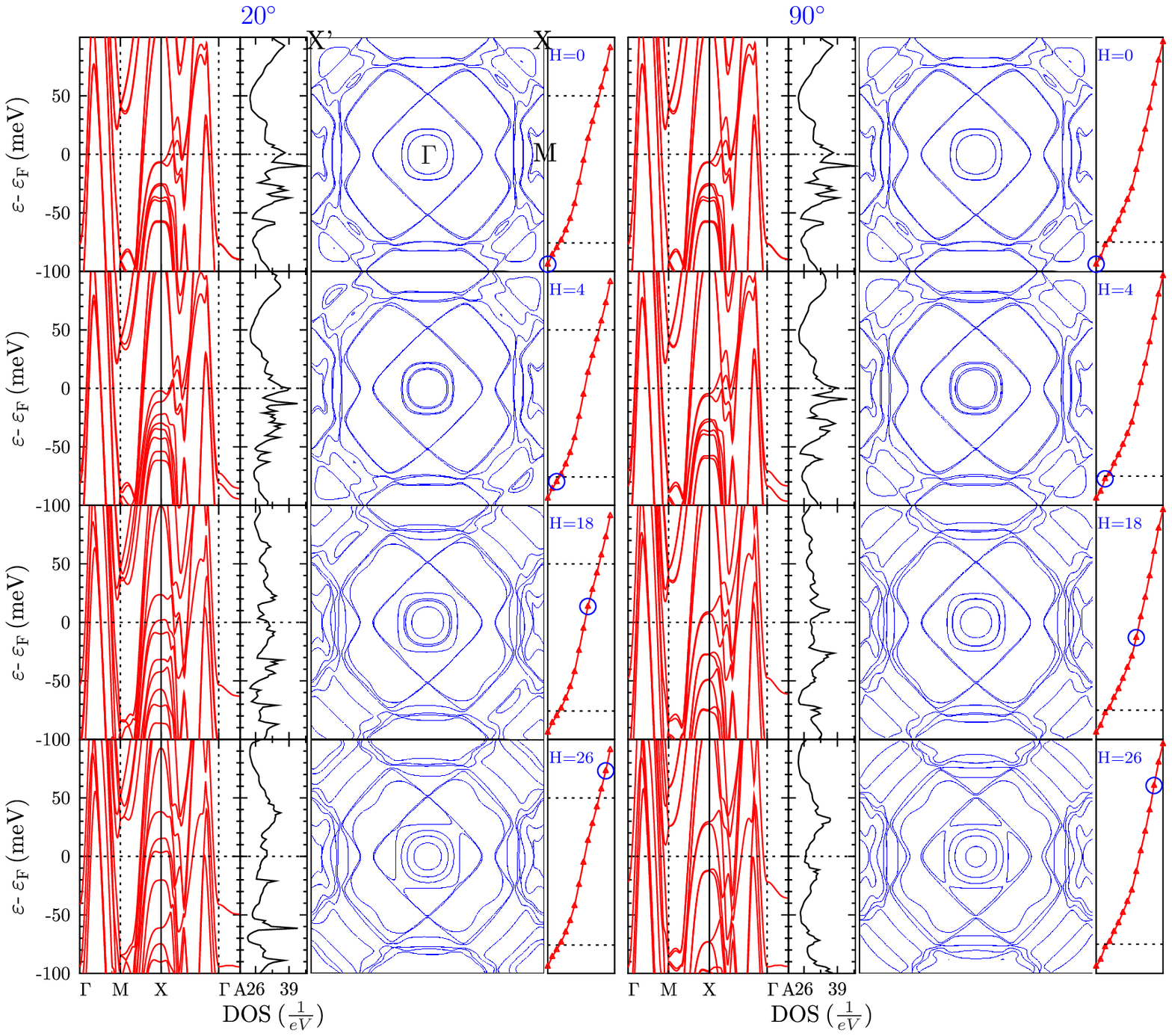}
\caption{(Color online) As Fig.~\ref{fig:bands-field-z}, here for
$\theta$=20$^{\circ}$ (left) and $\theta$=90$^{\circ}$ (right).
\label{fig:bands-field-20x}}
\twocolumngrid
\end{figure*}
After a near linear rise for $H$$>$5, close to $H$=16 a sudden change of slope for $M(H)$
takes place (with a possible signature in the free energy) and the magnetic moment
continues again nearly linearly. The upper first-order transition is not visible
anymore in the free-energy curve. This observation shows clear resemblance to the 
experimental findings of singular behavior for in-plane magnetic field compared to
strong out-of-plane $H$.~\cite{mac12} However from the present computations we can
not draw a definite conclusion concerning the shift of the phase boundaries with 
respect to $H$ and $\theta$. 
We are nevertheless in the position to shed more light on the angular-dependent
differences by showing in Fig.~\ref{fig:orbdep20x} the orbital contributions to the 
occupation and the magnetic moment for $\theta$=20$^{\circ}$, 90$^{\circ}$. For the
smaller $\theta$ the orbital-resolved differences compared to $H$ along $c$ are 
marginal, with mainly an obvious occupation difference between $d_{xz}$, $d_{yz}$ 
(with $n_{xz}$$>$$n_{yz}$) for $H$$>$22 which was absent before. 
Comparing the results for $H$ along $x$ to the case along $z$ shows again clear 
signature. The stronger occupation of $d_{xz}$ occurs at already small magnetic field 
and becomes substantial at large $H$. Furthermore the former area of diamagnetic 
response from $d_{xz}$, $d_{yz}$ has nearly vanished, both orbital responses are 
nearly indistinguishable from zero within the accuracy for small $H$.
Yet at least for $H$$>$15 the PM response with different
amplitude is clearly observable. Thus the dia/para competition in $d_{xz}$, $d_{yz}$
that seemed to be crucial for the MM behavior for $H$ along $z$ is nearly absent,
explaining the qualitative difference between in- and out-of-plane field. Coming back
to the phase-region shift with field and angle, and fixing such a shift to the dia/para
crossing for $d_{xz}$, $d_{yz}$, one might observe that this crossing indeed shifts
to the left, i.e. towards smaller $H$. However again the resolution is not accurate 
enough to render a unique statement concerning that question.
The impact of the finite angle $\theta$ shows also up in the changes of the QP states
with $H$, as documented in Fig.~\ref{fig:bands-field-20x}. While again for 
$\theta$=20$^{\circ}$ there are no major qualitative differences compared to the
case of $H$ along $c$ (e.g. also here the three low-lying bands at $X$ cross the
Fermi level with field), the pure in-plane field leads to clear modifications. Namely,
the two originally nearly degenerate lowest-energy bands at $X$ do {\sl not} split
with $H$, but cross $\varepsilon_{\rm F}$ together. In line with this, the four
$\gamma_2$ pockets in the BZ behave coherently for all field strengths and also the
$\delta$-$\alpha_1$ hybridization at large $H$ occurs now in a fourfold manner.
Importantly the third low-energy band (now also still degenerate with the fourth one)
remains below the Fermi level within the range of the studied magnetic field.
Thus $H$ along the $x$-axis leads to an avoided lifting of degeneracies, resulting in 
qualitative different magnetic response. However note that this on the other hand 
does not imply that the symmetry between $d_{xz}$, $d_{yz}$ is enforced, since the 
local orbital discrepancy is strongly {\sl increased} for the sole in-plane field 
(compare Fig.~\ref{fig:orbdep20x}).
\begin{figure}[t]
\centering
\includegraphics*[width=8.5cm]{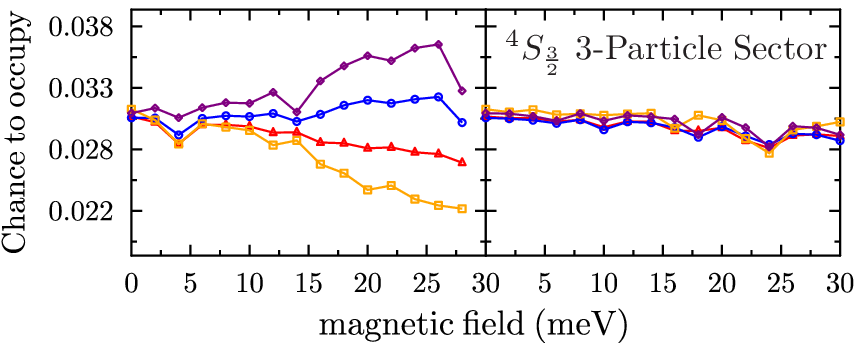}
\caption{(Color online) Splitting of the $^4S_{\nicefrac{3}{2}}$ multiplet 
for $\theta$=20$^{\circ}$ (left) and deferred splitting for 
$\theta$=90$^{\circ}$ (right).
\label{fig:multiplet-20x}}
\includegraphics*[width=8cm]{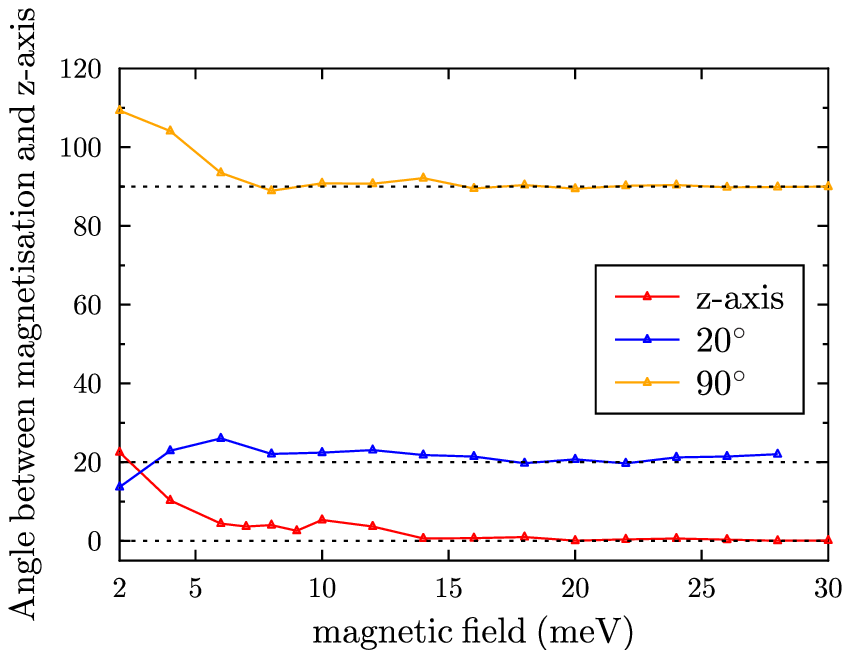}
\caption{(Color online) Angle between applied magnetic field $H$ and resulting
net magnetic Ru moment for  $\theta$=0, 20$^{\circ}$, 90$^{\circ}$.
\label{fig:angles}}
\end{figure}
Such selection-rule constraints depending on the magnetic-field direction show up 
also prominently when it comes to local-multiplet splittings, as shown in 
Fig.~\ref{fig:multiplet-20x} for our example of the $L$=0 $^4S_{\nicefrac{3}{2}}$ 
many-body state from the three-particle sector. While the $J_z$ splitting for 
$\theta$=20$^{\circ}$ remains vital, in the case of $H$ along $x$ the splitting is 
mostly absent. 
For further clarification, Figure~\ref{fig:angles} depicts the 
resulting angle between applied field and net magnetization, rendering it clear that 
only at large enough field strength the moment aligns along $H$. In this respect it 
appears as if for ${\bf H}$$\perp$${\bf c}$ the moment is somewhat more easily forced 
into the field direction.

\section{Summary and Discussion\label{sec:dis}}
The physical content of this work is twofold. First the interplay of spin-orbit
coupling and local Coulomb interactions was studied at equilibrium for the $n$=1,2 
layered strontium ruthenium oxides belonging to the Ruddlesden-Popper family. Thereby
we started from the realistic low-energy Kohn-Sham dispersion as obtained from 
Wannier-downfolding the bands from state-of-the-art LDA calculations. Then notably the
multi-orbital many-body effects (and its interlinking with spin-orbit effects) were 
treated beyond simple Hartree-Fock mean-field by utilizing proper self-consistent 
renormalizations due to strong correlations within rotationally invariant slave-boson 
theory at saddle-point. 

For both compounds it became evident that SOC and strong correlations together are 
important to account for the detailed low-energy electronic structure at small 
temperatures. Already in Sr$_2$RuO$_4$ the local Coulomb correlations effectively 
renormalized the spin-orbit interaction, leading to enhanced band-splittings for the 
$t_{2g}$ manifold close to the Fermi level. Its is therefore surely expected that the 
unconventional superconductivity at low $T$ has to be 
addressed by treating these both interaction types on equal footing.~\cite{dei11}
For the bilayer compound Sr$_3$Ru$_2$O$_7$ such an approach was shown to be essential
in order to describe the intriguing low-energy quasiparticle band structure and
density of states in close resemblance to existing ARPES and specific-heat 
measurements.~\cite{tam08,ros10} The renormalized spin-orbit
split bands give rise to extremely small energy scales, whereby close to the $X$ point
in the BZ an especially rich structure appears. However a straightforward decomposition
of the complicated renormalized band structure into distinct $d_{xy}$-, $d_{xz}$-, or 
$d_{yz}$-like bands seems difficult, the system looks like an intricate multi-orbital
system where subtle differences in the orbital contributions eventually play a 
crucial role.

Beyond the equilibrium study, a straightforward examination of the bilayer ruthenate
in applied magnetic field, based on the complete multi-orbital $t_{2g}$ Hamiltonian 
including
the Zeeman term in the presence of SOC, was presented. Depending on the magnetic-field
direction, metamagnetic transitions in line with first-order phase transitions were
verified. Taking a local viewpoint in that metallic system, the competition between 
paramagnetic ($d_{xy}$) and diamagnetic ($d_{xz}$, $d_{yz}$) contributions appears
to play a significant role for the MM phenomena. Moreover an orbital charge-transfer 
from $d_{xy}$ to $d_{xz}$, $d_{yz}$ with increasing $H$ was observed, though the 
orbital fillings between the latter two $t_{2g}$ orbitals only seem to deviate with 
larger field angle $\theta$. Concerning the itinerant QP states it became evident that
Lifshitz transitions close to the $X$ point (i.e. around the $\gamma_2$ pocket) may
partly be blamed for changes in the free energy and accompanying Fermi-surface 
reconstructions across the MM transitions. But moreover a substantial change of the
$\alpha_2$ sheet and an additional $\alpha_1$-$\delta$ hybridization is seen with
larger magnetic field. While the region around $X$ corresponds to propagation along 
the Ru-O-Ru bond with strong short-range variation, the latter inner sheets account
for rather isotropic in-plane transport with only long-range variation in the respective QP
wave function. Thus the physics of the MM transitions involves directional, short-range
processes at smaller $H$ and incorporates long-range mechanisms at
larger $H$. In addition, symmetry changes seem to take place between certain regions 
of the BZ. Albeit the overall $C_2^z$ symmetry of the ${\sl total}$ FS remains stable
with $H$, the original $C_4^z$-like symmetry between the $\gamma_2$ pockets is 
disturbed towards $C_2^z$ in the central MM phase region II, if 
${\bf H}$$||$${\bf c}$ holds. However that symmetry-change is absent for 
${\bf H}$$\perp$${\bf c}$ and a well-defined bounded MM region can not be identified.
Thus the present calculations reveal the qualitative differences between out-of-plane
and in-plane magnetic-field in accordance with experiment (see 
Ref.~\onlinecite{mac12} for a recent review). 

There remain however open questions. For instance, we may not draw definite conclusions
on the shifting of the MM transitions with field angle, whereas in some model 
studies~\cite{rag09,fis10} the shift of the MM phase region to lower magnetic fields  
with $\theta$ in line with experimental work was verified. Besides several other
possible reasons, this angle-dependent behavior might also be sensitive to the specific
choice for the magnitude of the interaction parameters $U$, $J_{\rm H}$ and $\lambda$ 
within the local Hamiltonian, also in conjunction with the nesting properties between the
various spin-polarized Fermi sheets.~\cite{fis10} 

Another important point concerns the appearance of nematic order, revealed in transport
studies for Sr$_3$Ru$_2$O$_7$ to escort the MM region.~\cite{bor07} We evidently see
symmetry changes in the FS geometry and standard representations of nematic order
parameters for a certain angular-momentum ordering channel $l$ of the form 
${\cal N}_l$=$\sum_{\bk}n(\bk)\,{\rm exp}[il\varphi(\bk)]$ (see e.g. 
Ref.~\onlinecite{fra10} for a review) display non-trivial behavior with $H$ depending
on $l$ and on the number of included bands. However the computed data does not exhibit 
convincing evidence for a well-defined quantification of nematicity along that
definition. Note that we also did not incorporate such an symmetry-breaking (i.e. 
forward-scattering) term explicitly in the Hamiltonian, as done in some model 
studies,~\cite{kee05,pue07,rag09,leewu09_2,fis10} and it might therefore be possible 
that we miss additional (energetically favorable) symmetry-breakings in our mean-field 
approach. For the present problem, the nematic order parameter has frequently also 
been defined via the filling difference between the
$d_{xz}$, $d_{yz}$ orbitals.~\cite{rag09,leewu09_2} Yet this definiton seems dangerous,
since the $C_2^z$ symmetry of Sr$_3$Ru$_2$O$_7$ is broken already by the
equilibrium crystal structure and LDA calculations reveal such a nominal filling 
difference for $H$=0. On the other hand our extended-LDA calculations point towards 
an initial alignment of these sub-orbital fillings with magnetic field. Only for
larger ($H$, $\theta$) a true $d_{xz}$, $d_{yz}$ filling differentiation occurs.
In this respect its also noteworthy that here the magnetic-field tilting towards the
$ab$-plane takes place inbetween the Ru-O-Ru bond, i.e. along the diagonal $a$-axis
of the in-plane square lattice. It would therefore be interesting to check further 
additional in-plane directions, especially along Ru-O-Ru.

This brings us to possible extensions of the current work on Sr$_3$Ru$_2$O$_7$. For 
present numerical reasons, the different Ru ions in the bilayer unit cell were 
assumed equivalent by symmetry (as true for the equilibrium crystal structure). But we
easily expect that the MM phase regions would generally benefit from such a
symmetry breaking within the unit cell. If orbital-liquid scenarios or the physics of 
domain structures~\cite{sti11,mac12} in the compound are vital, one has to come up with
even more sophisticated real-space picturings. Inter-site Coulomb interactions are
here neglected, but may also be a source for the observed symmetry 
breakings.~\cite{pue10} Furthermore since this work was performed by using a 
postprocessing scheme to existing LDA calculations, elaborating on a complete 
charge-selfconsistent approach with the proper feedback of the electronic self-energy 
onto the Kohn-Sham charge density could surely enhance the MM response. Accounting
for finite-temperature effects is an additional further important aspect in order
to reveal the intricate thermodynamics of the MM region.~\cite{ros11} Last but not
least, at the moment the method relies on mean-field theory. Since in other strongly
correlated materials it is already found that intricate self-energy effects especially take
place close to van-Hove singularities,~\cite{pei11} local quantum fluctuations should
be included in future Sr$_3$Ru$_2$O$_7$ studies. Non-local quantum spin fluctuations may have 
relevant impact on the low-energy physics, since the material is prone to magnetic order. 
Nonetheless the present realistic formalism yields promising results and shows that 
extended-LDA calculations are in principle capable of addressing the challenging low-energy 
physics of the layered ruthenates.

\begin{acknowledgments}
We thank D.~Grieger for helpful discussions.
Financial support from the Free and Hanseatic City of Hamburg in the context of the 
Landesexzellenzinitiative Hamburg as well as the DFG-FOR 1346 is gratefully 
acknowledged. Computations were performed at the local computing center of the 
University of Hamburg as well as the North-German Supercomputing Alliance (HLRN) 
under the grant hhp00026.
\end{acknowledgments}

\bibliographystyle{apsrev4-1}
\bibliography{bibextra}
\end{document}